\documentclass[aps,prb,twocolumn, superscriptaddress]{revtex4-1}
\usepackage{graphicx}
\usepackage{array}
\usepackage{dcolumn}
\usepackage{bm}
\usepackage{float}
\usepackage{enumitem}
\usepackage[colorinlistoftodos, textwidth=4cm, shadow]{todonotes}

\begin{document}


\title{Magnetic structure and phase stability of the van der Waals bonded ferromagnet Fe$_{3-x}$GeTe$_2$}

\author{Andrew F. May}
\email{mayaf@ornl.gov}
\affiliation{Materials Science and Technology Division, Oak Ridge National Laboratory, Oak Ridge, TN 37831}
\author{Stuart Calder}
\affiliation{Quantum Condensed Matter Division, Oak Ridge National Laboratory, Oak Ridge, TN 37831}
\author{Claudia Cantoni}
\affiliation{Materials Science and Technology Division, Oak Ridge National Laboratory, Oak Ridge, TN 37831}
\author{Huibo Cao}
\affiliation{Quantum Condensed Matter Division, Oak Ridge National Laboratory, Oak Ridge, TN 37831}
\author{Michael A. McGuire}
\affiliation{Materials Science and Technology Division, Oak Ridge National Laboratory, Oak Ridge, TN 37831}

\date{\today}

\begin{abstract}
The magnetic structure and phase diagram of the layered ferromagnetic compound Fe$_3$GeTe$_2$ has been investigated by a combination of synthesis, x-ray and neutron diffraction, high resolution microscopy, and magnetization measurements.  Single crystals were synthesized by self-flux reactions, and single crystal neutron diffraction finds ferromagnetic order with moments of 1.11(5)$\mu_B$/Fe aligned along the $c$-axis at 4\,K.  These flux-grown crystals have a lower Curie temperature $T_{\textrm{c}}\approx$150\,K compared to crystals previously grown by vapor transport ($T_{\textrm{c}}$=220\,K).  The difference is a reduced Fe content in the flux grown crystals, as illustrated by the behavior observed in a series of polycrystalline samples.  As Fe-content decreases, so does the Curie temperature, magnetic anisotropy, and net magnetization.  In addition, Hall effect and thermoelectric measurements on flux-grown crystals suggest multiple carrier types contribute to electrical transport in Fe$_{3-x}$GeTe$_2$ and structurally-similar Ni$_{3-x}$GeTe$_2$.
\end{abstract}

\maketitle

\section{Introduction}

The extensive research on graphene and ultra-thin transition metal dichalcogenides (TMD) is naturally progressing into studies of van der Waals (VDW) bonded heterostructures and application-oriented configurations.\cite{Geim2013,TMDC_NatChem,RecentBeyondGraphene}  VDW heterostructures are produced from layer-by-layer stacking of monolayer (or few-layer) materials, which can be obtained from exfoliation of bulk sources.\cite{BeyondGraphene}  These engineered structures will inevitably lead to unique properties and potentially new technologies, in part due to the ability to combine materials with complementary functionalities.  This was recently demonstrated with the realization of highly-efficient, picosecond photoresponse in a heterostructure of graphene (fast response) and WSe$_2$ (high efficiency).\cite{Massicotte2015}   The construction of VDW heterostructures using topological insulators (TI) will certainly provide ample phase space for exploring novel states of matter and theoretical predictions.  Currently, there is interest in controlling the dispersion and carrier type of the Dirac surface states of a TI using van der Waals heterostructures that allow separate tuning of bulk and surface states.\cite{Chang2015}

VDW-bonded ferromagnets are of interest as building blocks for heterostructures designed for use in spin-based information technologies, either for the direct exploitation of their magnetic properties or via magnetic proximity effects.  The latter permits the use of nominally non-magnetic materials in spintronics, and is being pursued using EuO/graphene heterostructures.\cite{Swartz2012,Yan2013}  Similarly, skyrmions are of interest from a fundamental perspective and for their ability to potentially enable low-power spintronics, and these spin states are stabilized by a reduction from three to two dimensions and by the presence of Rashba spin-orbit coupling.\cite{Yu2011,Banerjee2014}  In general, spin-orbit coupling within heterostructures should yield interesting spin structures and magnetoelectric transport.

Of the `next-generation' VDW materials, CrI$_3$ and CrSiTe$_3$-type compounds have recently been identified as promising systems with the potential for long-range magnetism in monolayers.\cite{Casto2015,Sivadas2015,William2015,McGuire2015,Zhang2015}   These materials are ferromagnetic at approximately 61\,K and 33\,K, respectively,\cite{McGuire2015,Carteaux1995} which is somewhat low for spintronic applications.  Interestingly, magnetic odering temperatures have been predicted to increase when monolayers are constructed,\cite{Sivadas2015} and initial experimental results seem to confirm this behavior in CrSiTe$_3$.\cite{MingWei2015}  Considering this, the VDW-bonded compound Fe$_3$GeTe$_2$ may be of particular interest because the bulk is ferromagnetic near 230\,K.\cite{Deiseroth2006}

\begin{figure}[hb!]%
\includegraphics[width=0.9\columnwidth]{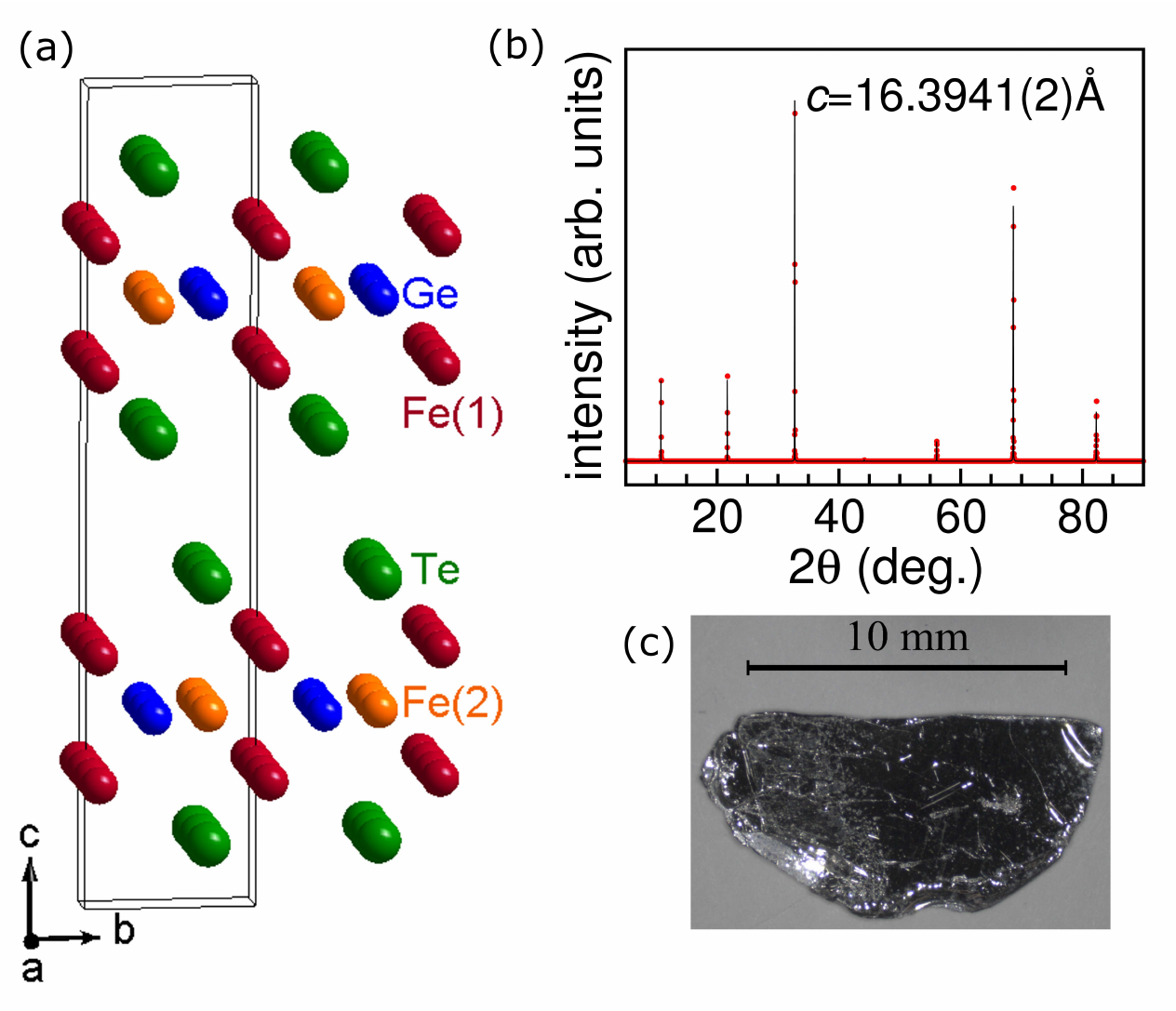}%
\caption{(color online). (a) Image of crystal structure with atomic positions labeled; Fe(1)-Fe(1) bonds pierce the center of each hexagon formed by Fe(2)-Ge (space group $P6_3/mmc$ - no. 194).  (b) X-ray diffraction data for an as-grown facet of Fe$_{3}$GeTe$_2$ showing the $c$-axis normal with exclusively 00l diffraction peaks observed; a Le Bail fit (solid line) yielded the $c$-axis lattice parameter shown in the image. (c) A picture of a large single crystal.}%
\label{00l}%
\end{figure}

Fe$_{3}$GeTe$_2$ contains Fe$_3$Ge slabs separated by VDW-bonded Te layers, as shown in Fig.\,\ref{00l}(a).  The Fe$_3$Ge slabs contain Fe(1)-Fe(1) pairs across a hexagonal network built by Fe(2)-Ge, and this is structurally similar to the more three-dimensional Fe$_{2-x}$Ge compounds.\cite{Kanematsu1965}  Fe$_{3}$GeTe$_2$ is an itinerant ferromagnet, with Curie temperatures of 220 and 230\,K reported, and an estimated spontaneous magnetization of 1.6$\mu_B$/Fe at 0\,K.\cite{Deiseroth2006,Chen2013}  Previous anisotropic magnetization measurements on crystals grown using chemical vapor transport suggested that the $c$-axis is the easy axis, and an anisotropy field of at least 5\,T was demonstrated at 10\,K.\cite{Chen2013}  The Fe(2) position was reported to have a small concentration of vacancies (17\%), but chemical characterization suggested the composition is Fe$_3$GeTe$_2$.\cite{Deiseroth2006} Interestingly, Fe vacancies in Fe$_{2-x}$Ge are also concentrated on the chemically similar Fe(2) position and a wide, complex phase-width is observed.\cite{Kanematsu1965}  A modest phase-width is thus expected in Fe$_3$GeTe$_2$ based on the published crystallographic data, its intermetallic nature, and structural similarities with Fe$_{2-x}$Ge compounds. Upon completion of this work, the existence of a phase-width was independently confirmed, though the influence on the structure and magnetic properties was not reported.\cite{Verchenko2015}

We report the growth of single crystals via a molten flux technique, together with a study of polycrystalline samples that confirms a phase width and corresponding response in the magnetic properties.  We find that the lattice parameters, Curie temperature and saturation magnetization vary smoothly with Fe concentration, though the $a$ and $c$ lattice parameters trend oppositely.  Reduced Fe content in the flux grown crystals is found to be responsible for their lower $T_{\textrm{c}}$ when compared to prior reports.  Additionally, we find that Fe$_{3-x}$GeTe$_2$ itinerant ferromagnets possess multi-carrier electronic transport.

\section{Experimental Details}

Single crystals were grown from Fe-Ge-Te `self'-fluxes, and a melt composition of Fe$_2$GeTe$_4$ produced the largest crystals of those investigated.  Crystals were also successfully grown from the compositions FeGe$_2$Te$_4$ and FeGeTe$_2$ using similar heating procedures, though these crystals were typically smaller and had slightly lower Curie temperatures. High purity elements from Alfa Aesar (Fe 99.98\%, Ge 99.9999\%, and Te 99.9999\%)  were combined in Al$_2$O$_3$ crucibles and sealed in evacuated quartz ampoules.  A crucible filled with quartz wool was placed on top of the growth crucible to catch the excess flux during centrifugation.  The melt was homogenized at 950$^{\circ}$C for approximately 12\,h, then cooled slowly to 675$^{\circ}$C, at which temperature the ampoules were removed from the furnace and placed in a centrifuge to expel the excess flux. A variety of cooling rates were found to produce crystals, and in this paper we report data from crystals obtained after cooling at 1 and 3$^{\circ}$/h (the properties were observed to be equivalent).  We also synthesized crystals of Ni$_3$GeTe$_2$ to provide a non-magnetic reference material during our investigation of transport properties.  Crystals of Ni$_3$GeTe$_2$ were grown from a melt of NiGeTe$_2$ cooled at 3$^{\circ}$/h, with the flux removed at 660$^{\circ}$C.  The crystals of Ni$_{3}$GeTe$_2$ were generally smaller than those of Fe$_{3}$GeTe$_2$.  Upon characterization, both were found to be transition metal deficient.

Polycrystalline samples with nominal Fe concentrations between Fe$_{3.10}$GeTe$_2$ and Fe$_{2.60}$GeTe$_2$ were prepared by grinding Fe powder, Ge powder, and Te shot in a He filled glove box.  The mixture was transferred, under He, to a vacuum line where the quartz ampoule was sealed under vacuum.  The samples were heated at 675$^{\circ}$C for approximately 10\,d.  The as-reacted Fe$_{3-x}$GeTe$_2$ samples were found to be either slightly sintered and dull black for small $x$, or heavily sintered with visible grain growth/crystallization for large $x$.  At this point, the samples were analyzed with x-ray diffraction and magnetization measurements.  Small pellets were then fired briefly at 600$^{\circ}$C to facilitate isothermal magnetization and energy dispersive spectroscopy (EDS) measurements.  EDS measurements were performed using a Bruker Quantax 70 EDS system on a Hitachi TM-3000 microscope.

Crystals from the flux-grown reactions are much larger than those needed for single crystal x-ray diffraction.  Due to the ease with which the crystals cleave and deform, the process of cutting or crushing the crystals induces significant damage. Therefore, a small crystal from the polycrystalline reaction with nominal composition Fe$_{2.75}$GeTe$_2$ was selected for single crystal x-ray diffraction.  Data were collected at 173\,K on a Bruker SMART APEX CCD, using Mo-$K\alpha$ radiation ($\lambda$ = 0.71073\,\AA, graphite monochromator).  For refinement of the crystal structure, absorption corrections were applied with SADABS and SHELXL-97 was used to refine the data, and the atomic coordinates were standardized with Structure Tidy within PLATON.\cite{Shelxl97,StructureTidy,Platon}  The refinement utilized 138 unique reflections from 1885 reflections and 12 refinement parameters.  The Goodness of Fit was 1.303 while R$_{\mathrm{int}}$ = 0.0280.  Powder x-ray diffraction data were collected using a PANalytical X'Pert Pro MPD with a Cu K$_{\alpha,1}$ ($\lambda$=1.5406\,\AA) incident beam monochromator, and Le Bail and Rietveld refinements were performed using FullProf.\cite{FullProf}

Magnetization measurements were performed in a Quantum Design Magnetic Property Measurement System, as well as with the AC Magnetic Susceptibility Option on a Quantum Design Physical Property Measurement System (PPMS).  A PPMS was also used to measure the Seebeck coefficient, thermal conductivity and electrical resistivity using the Thermal Transport Option.  For this measurement, gold-coated copper leads were attached to the crystals using H20E Epo-Tek silver epoxy.  Hall effect measurements and magnetoresistance measurements were performed using a standard four-point configuration with Pt wires attached via silver paint (DuPont 4929N).  The Hall resistance $\rho_{\mathrm{H}}$ was obtained via the odd-in-\textit{\textbf{H}} part of the transverse resistance $\rho_{\mathrm{H}}=(\rho_{xy}[\textit{\textbf{H}}]-\rho_{xy}[\textit{\textbf{-H}}]$)/2 with maximum magnetic fields of magnitude $H$=80\,kOe applied along the $c$-axis.  Magnetoresistance was obtained from the even-in-\textit{\textbf{H}} portion of the longitudinal resistance with fields applied along the $c$-axis.  At 2\,K and 80\,kOe, the magnetoresistance was less than 2\% and we have excluded the data from the manuscript.

To probe the microscopic crystal and magnetic structure, we performed single crystal neutron diffraction on the Four-Circle Diffractometer (HB-3A) at the High Flux Isotope Reactor (HFIR), ORNL. A single crystal of approximate dimensions 4\,mm\,$\times$\,4\,mm was mounted on an Al rod inside a CCR. Using an incident wavelength of 1.003\,$\rm\AA$ (Si(331) monochromator), measurements were performed between 4\,K and room temperature. A large number of reflections were collected at 220\,K and 4\,K to determine the nuclear and magnetic structures, respectively.  Neutron powder diffraction was performed on the HB-2A Neutron Powder Diffractometer at HFIR, ORNL. A wavelength of 2.41\,$\rm\AA$ was used in all measurements, and this was selected with a Ge(113) monochromator; the samples were placed in Al cans.  All neutron diffraction data were refined using the program FullProf.\cite{FullProf}

High resolution scanning transmission electron microscopy (STEM) was performed using a Nion UltraSTEM200 microscope operating at 200\,keV, equipped with a Gatan Enfinium spectrometer for the in situ collection of electron energy loss spectra (EELS).  Samples were examined in both plan view and cross-sectional orientations.  Samples were prepared by a combination of polishing and Ar$^+$ ion milling using a voltage of 2\,keV. Contact with moisture was avoided during sample preparation.

\section{Results and Discussion}

The single crystals obtained from self-flux growths are thin plates with dimensions reaching greater than 1\,cm (see Fig.\,\ref{00l}).  X-ray diffraction data collected from the surface of as-grown facets confirm the expected orientation with [00l] normal to the facet, as shown in Fig.\,\ref{00l}.  A Le Bail fit to the diffraction data in Fig.\,\ref{00l} yielded $c$=16.3941(2)\AA. Rietveld refinement of data collected on ground crystals resulted in $a$=3.9536(7)\AA\, and $c$=16.396(2)\AA.  These values differ sharply from those in the literature, where $a$=3.9910(10)\AA\, and $c$=16.336(3)\AA\, were reported from room temperature single crystal x-ray diffraction.\cite{Deiseroth2006}

\begin{figure}[h]%
\includegraphics[width=0.9\columnwidth]{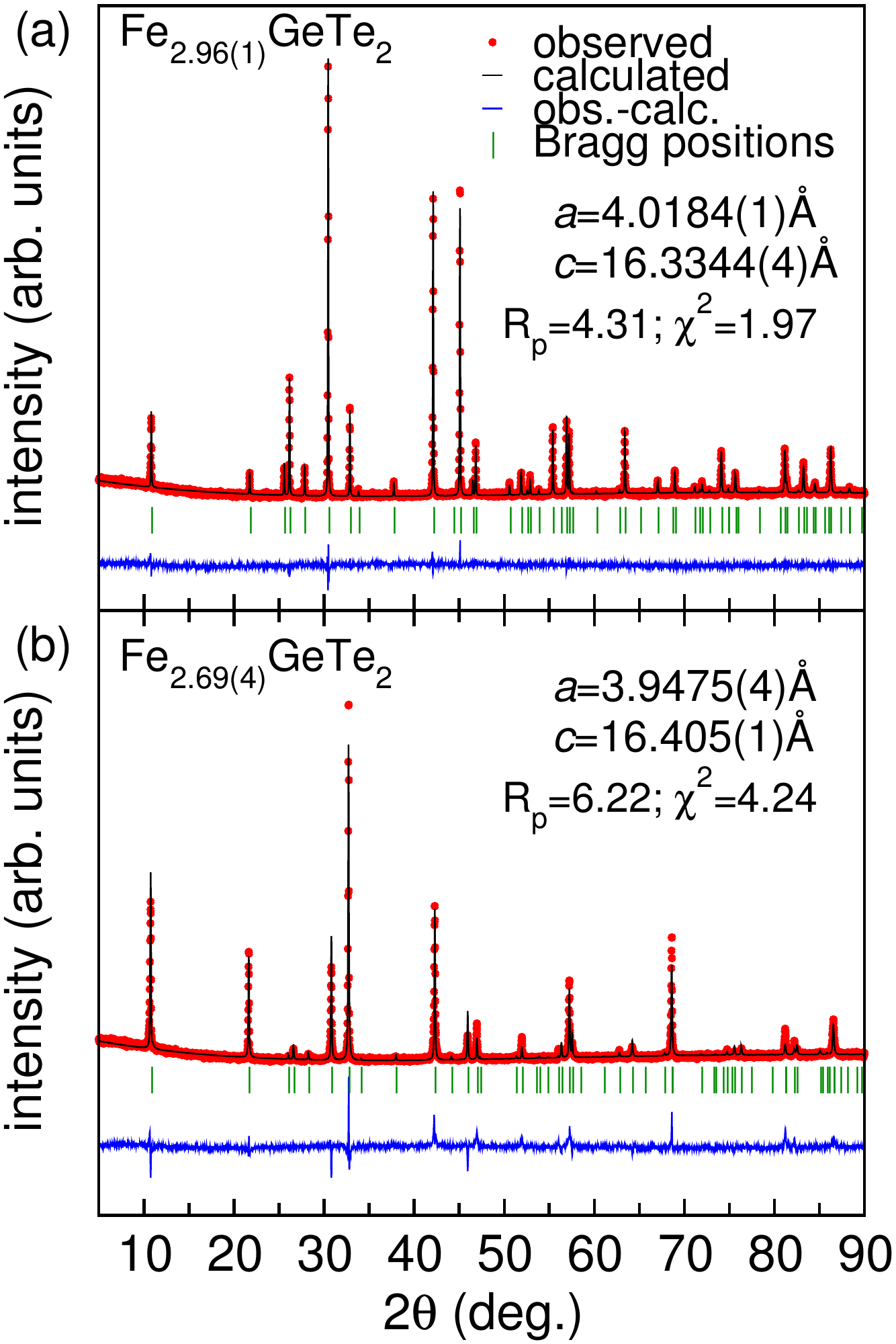}%
\caption{(color online). Powder x-ray diffraction data for two polycrystalline samples with refined lattice parameters and compositions provided.}%
\label{poly_xrd}%
\end{figure}

Ni$_{3}$GeTe$_2$ crystals were grown from a self-flux to provide a non-magnetic reference material during the characterization of physical properties.  The lattice parameters obtained for our Ni$_{3}$GeTe$_2$ crystals also differ from those in the literature, with refinement of powder diffraction data yielding $a$=3.8373(2)\AA\, and $c$=16.048(2)\AA.  This compares to $a$=3.9110(10)\AA\, and $c$=16.022(3)\AA\, previously reported for crystals with a composition of Ni$_{2.95}$GeTe$_2$ obtained from refinement of single crystal x-ray diffraction data.\cite{Deiseroth2006}  Thus, for both Fe$_{3}$GeTe$_2$ and Ni$_{3}$GeTe$_2$ grown from a self-flux, the $a$ lattice parameter is smaller and the $c$ lattice parameter is slightly larger than the literature reports for crystals obtained from nominal 3-1-2 compositions (formed via solid state reactions).

In addition to having different lattice parameters than those in the literature, the flux-grown Fe$_3$GeTe$_2$ crystals have lower Curie temperatures than those previously reported for Fe$_3$GeTe$_2$.  The crystals were grown from self-fluxes, which minimize the chance for extrinsic doping.  We thus speculated that Fe$_3$GeTe$_2$ contains a non-trivial phase width that influences structure and physical properties.  Our EDS results suggest Fe and Ge deficiencies may exist, with a composition of Fe$_{2.91(3)}$Ge$_{0.95(4)}$Te$_{2.00(4)}$ obtained.  For the Ni-based analogue, EDS yielded Ni$_{2.40(4)}$Ge$_{1.01(3)}$Te$_{2.00(3)}$ for crystals grown from NiGeTe$_2$.

To investigate the phase width of Fe$_{3-x}$GeTe$_2$, polycrystalline samples were produced via solid-state reactions with nominal compositions Fe$_{3-x}$GeTe$_2$ with  $-0.1 \le x \le 0.4$.  These materials were characterized using powder x-ray diffraction, EDS, and magnetization measurements.  We begin by presenting the results of our crystallographic studies, and then return to the magnetic properties and detailed characterization of our single crystals.  When necessary for unit conversion, we have utilized compositions obtained by Rietveld refinement.  When discussing single crystals, we present data for crystals grown from Fe$_2$GeTe$_4$ melts, and refer to these by the composition obtained from refinement of single crystal neutron diffraction data, Fe$_{2.76}$Ge$_{0.94}$Te$_2$.

\subsection{Structure and Composition}

All polycrystalline samples were found to be phase pure within the limits of our powder x-ray diffractometer, with the exception being a sample of nominal composition Fe$_{2.60}$GeTe$_2$.  This sample contained GeTe and FeTe$_2$ impurities, and data for this sample have been excluded from the manuscript.  Our results thus suggest the Fe-deficient phase boundary is likely reached near Fe$_{2.7}$GeTe$_2$.  We have not investigated the phase width with regard to Te or Ge.

The sample of nominal composition Fe$_{3.10}$GeTe$_2$ did not possess any obvious impurities by x-ray diffraction.  The refined composition for this sample is Fe$_{2.97(2)}$GeTe$_2$, while large-area (300\,$\mu$m diameter) EDS scans yielded Fe$_{3.07(2)}$Ge$_{0.92(2)}$Te$_{2.00(2)}$.  The EDS results are likely influenced by minor secondary phases that are not easily detected by x-ray diffraction.  An Fe-rich impurity was observed in back scattered electron images, and we were not able to isolate large grains of the main phase for EDS due to resolution limits. Therefore, we focus on compositions obtained from Rietveld refinements of the powder diffraction data. In the structurally-similar Ni$_3$GeTe$_2$ compound, occupation of an interlayer Ni(3) position at (0,0,0) was detected by single crystal x-ray diffraction, though occupation of this site was not observed in Fe$_3$GeTe$_2$.\cite{Deiseroth2006} We performed refinements including the hypothetical Fe(3), and the results suggest that an occupation of 3(1)\% may exist for the nominal Fe$_{3.10}$GeTe$_2$.  This seems reasonable, but the result is dependent on the range of data analyzed (in part due to a connection to the refinement of sample texture).  Based on our single crystal diffraction and electron microscopy, discussed below, we feel confident that there is not any significant occupation of this interlayer site in the nominally Fe-deficient samples (at least for $x$$\approx$0.25).  We therefore utilized the published crystal structure, with two Fe positions, to refine our diffraction data and report compositions based on this refinement.

\begin{figure}[ht]%
\includegraphics[width=0.9\columnwidth]{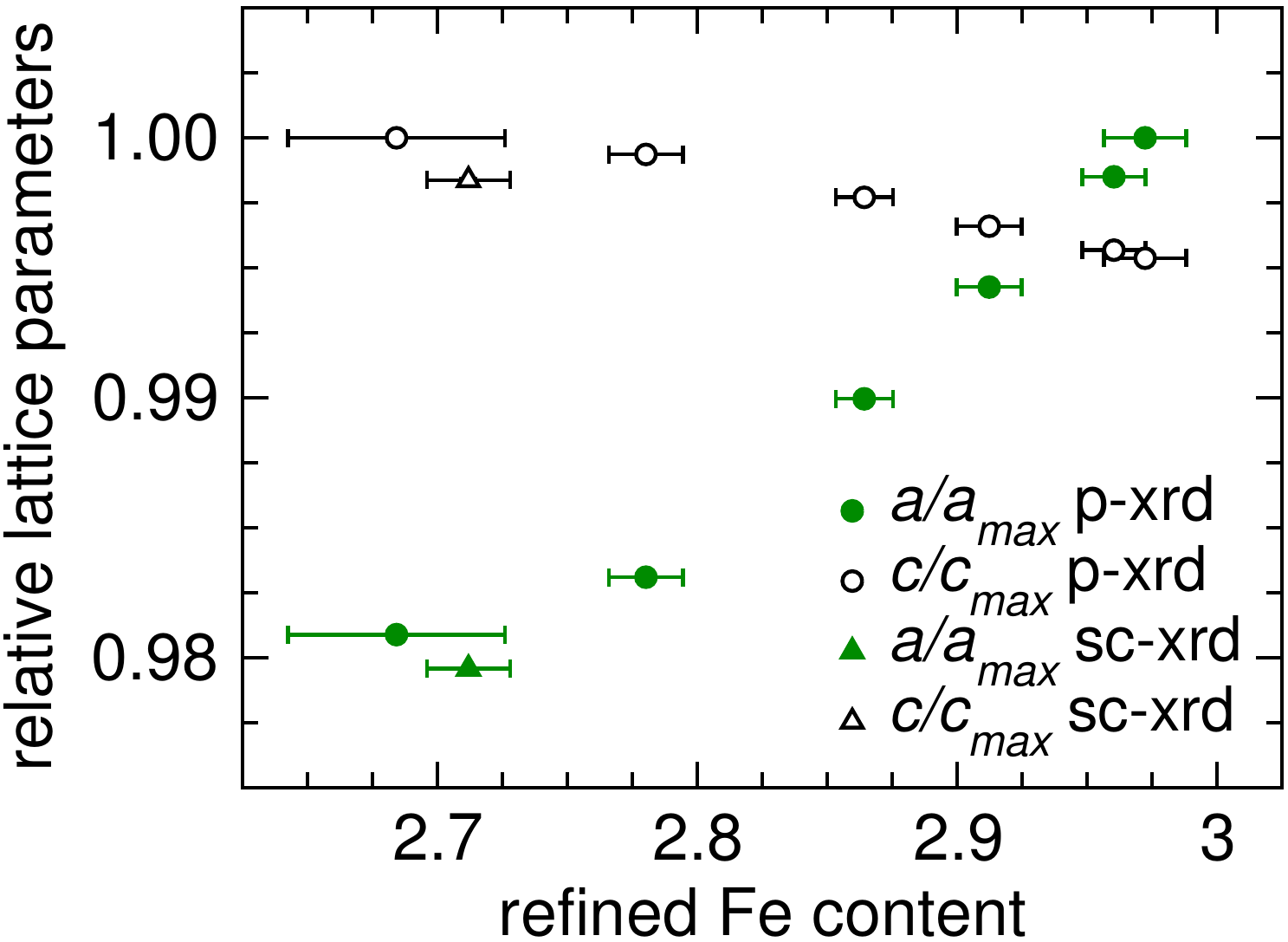} \\
\caption{Normalized lattice parameters as a function of refined Fe content for Fe$_{3-x}$GeTe$_2$ samples (powder x-ray diffraction `p-xrd' at room temperature), including results obtained from single crystal x-ray (`sc-xrd') diffraction data collected at 173\,K; $a_{max}$\,=\,4.0244(1)\,\AA and $c_{max}$\,=\,16.405(1)\,\AA.  Vertical error bars are smaller than the size of the data markers.} %
\label{occ}%
\end{figure}

Figure \ref{poly_xrd} presents powder x-ray diffraction data and Rietveld refinements for samples of nominal composition Fe$_{3.00}$GeTe$_2$ and Fe$_{2.75}$GeTe$_2$, which refined to the compositions Fe$_{2.96(1)}$GeTe$_2$ and Fe$_{2.69(4)}$GeTe$_2$, respectively. The lattice parameters obtained from refinement of the data in Fig.\,\ref{poly_xrd} confirm that the Fe-deficient sample has smaller $a$ and larger $c$ than the Fe-rich sample.  This trend is shown in Fig.\,\ref{occ} for all polycrystalline samples, where the normalized lattice parameters are plotted as a function of Fe content.  Data obtained from single-crystal x-ray diffraction (sc-xrd) are also included to confirm this behavior; note that the sc-xrd data were collected at 173\,K where slightly smaller lattice parameters are expected.  The increase in $a$ with increasing Fe may be expected, with the lattice expanding as Fe(2) vacancies are filled.  Indeed, a similar trend in $a$ occurs for Fe$_{2-x}$Ge in the range 0.15 $< x <$ 0.7, though $c$ was found to increase with increasing Fe content in this more 3D compound.\cite{Kanematsu1965}

Given the layered structure, we expect the greatest influence of the increased occupation of Fe(2) to be observed within the (Fe$_3$Ge) planes.  Indeed, the relative expansion of $a$ is greater than the relative contraction of $c$ (see Fig.\,\ref{occ}).  We therefore expect the changes in $c$ to be driven by the changes in $a$.  That is, the decrease in $c$ is due to an expansion of the basal plane that allows Te atoms to get closer to the Fe$_3$Ge layer, while increased occupation of Fe(2) also leads to greater Fe(2)-Te bonding. Consistent with this, Fe(2)-Te bond distances are smallest in the Fe-rich samples, as are Fe(1)-Fe(1) bond distances.  As a result of Te being pulled towards the Fe$_3$Ge layers, the structure collapses along $c$ to maintain the van der Waals bonds (the length of which increases with increasing Fe).  If the hypothetical Fe(3) becomes occupied at high Fe concentrations, its presence could lead to increased bonding that would also reduce $c$.  However, refinements with Fe(3) reveal a larger Fe(3)-Te bond distance for higher Fe contents, suggesting that Fe(3) does not play a role in the contraction of $c$.

We note that the occupancy of Fe(2) is correlated to the refinement of texture (preferred orientation) in the powder diffraction data.  In these samples, the texture physically increases with decreasing Fe content.  This trend can be observed visually, with the Fe-deficient samples demonstrating more grain growth during the reaction.  Not refining the texture results in much larger residuals, and thus the preferred orientation was refined for all samples and care was taken to minimize the texture during sample preparation.  The main result is not strongly influenced by this, though, and the nominal composition clearly influences the amount of Fe in the final specimen and the trends are consistent.  Also, to simplify the refinement, we did not allow Ge content to vary and fixed an overall displacement parameter.

\begin{table}[ht!]
  \caption{Selected data from refinements of single crystal x-ray diffraction collected on a crystal from a reaction of nominal composition Fe$_{2.75}$GeTe$_2$. Atomic coordinates are Fe(1):  0,0,$z$; Fe(2): $\frac{1}{3}$,$\frac{2}{3}$,$\frac{1}{4}$; Te:  $\frac{1}{3}$,$\frac{2}{3}$,$z$; Ge: $\frac{1}{3}$,$\frac{2}{3}$,$\frac{3}{4}$.}
  \label{scxrd}
  \begin{tabular}{lcc}
   \hline
    \textit{a}  (\AA)         & 3.9421(9) \\
    \textit{c}   (\AA)        & 16.378(5) \\
    R$_1$ (all data)          & 0.0461   \\
    wR$_2$ (all data)         & 0.1019    \\
    Goodness of fit				  	& 1.306     \\
    Fe(1) $z$ coord.					& 0.1721(2)\\
    Te		$z$  coord.         & 0.0900(1) \\	
   \hline
  \end{tabular}
\end{table}

The potential existence of interlayer Fe is important beyond understanding the modifications to the lattice with changing Fe content.  As shown below, the total Fe content clearly influences the magnetic properties and interlayer Fe could play a particularly important role in determining the saturation magnetization and coercive field.  In regards to the chemistry and structure of these materials, the existence of interlayer Fe would imply a more three-dimensional material.  As such, interlayer Fe would likely impede the production of thin-layers by cleaving bulk crystals.  Similarly, if an amount of interlayer Fe can be controlled, it would likely influence the anisotropy of the electronic and magnetic properties.  Interestingly, as discussed below, the samples with the highest Fe content have the largest anisotropy in the magnetic properties.  While our powder diffraction results suggest Fe(3) is not occupied, we felt that additional investigation was warranted.

We investigated the issue of interlayer Fe with single crystal x-ray and neutron diffraction, and electron microscopy.  The refinement results from the x-ray diffraction data are shown in Table \ref{scxrd}, and we clearly observe the partial occupation of Fe(2) with a refined composition Fe$_{2.71(2)}$GeTe$_2$ (recall this is for a crystal obtained from the Fe$_{2.75}$GeTe$_2$ polycrystalline reaction; Ge vacancies were not observed). Electron density was not detected at an interlayer Fe(3) position, consistent with earlier work for crystals selected from a polycrystalline sample of nominal composition Fe$_{3}$GeTe$_2$.\cite{Deiseroth2006}  Our single crystal neutron diffraction on flux-grown crystals, discussed in more detail below, does not find any strong evidence for occupation of the Fe(3) position.

A large number of vacancies are refined on the Fe(2) position from both the neutron and x-ray single crystal diffraction data. When filled, these Fe(2) positions are bonded to three in-plane Ge atoms.  As vacancies are introduced, Fe(2)-Ge bonding decreases (on average) and this results in a large in-plane displacement parameter for Ge, as shown in Table \ref{tab:U}.  Consistent with this, we refine a larger concentration of vacancies on Fe(2) compared to Ref. \citenum{Deiseroth2006}, and our refined $U_{11}$  for Ge is also larger than previously reported.  However, the previous study did report a strong anisotropy for the displacement parameters of Ge, and a similar $U_{33}$ to that shown in Table \ref{tab:U} was reported.\cite{Deiseroth2006} In the binary compound Fe$_{1.60}$Ge$_2$, it has been suggested that the Ge atoms actually move off of their site and the symmetry is potentially broken.\cite{Malaman1980} 

The data in Table \ref{tab:U} correspond to refinement of single crystal x-ray diffraction data, where only Fe(2) was found to be partially occupied.  We obtained a fractional occupation of 0.71(2) for Fe(2), and this was utilizing a constraint to maintain equal displacement parameters for Fe(1) and Fe(2). When refined separately, the displacement parameters of Fe(2) are observed to be rather small within the basal plane ($U_{11}$=0.004(2)).  In this case, the refined occupancy of Fe(2)=0.69(2) is very similar to that obtained from the restrained fit.

\begin{table}
  \caption{Anisotropic displacement parameters for Fe$_{3-x}$GeTe$_2$ from refinements of single crystal x-ray diffraction data at $T$=173\,K; note $U_{11} = U_{22} = 2U_{12}$.}
  \label{tab:U}
  \begin{tabular}{lcccc}
    \hline
         species &  $U_{11}$ & $U_{33}$ & $U_{eq}$ & occupancy \\ 
      &  \AA$^2$ & \AA$^2$ & \AA$^2$ & fractional \\
    \hline
         Fe(1)  &  0.0069(12)  & 0.014(2)    & 0.0091(10)  & 1 \\
         Fe(2)  &  0.0069(12)  & 0.014(2)    &  0.0091(10) & 0.71(2) \\
         Ge     &  0.044(2)    &  0.015(2)   &  0.034(1)   & 1\\
         Te     &  0.0081(6)   &  0.0152(8)  &  0.0105(5)  & 1\\
   \hline
  \end{tabular}
\end{table}

We also performed high resolution scanning transmission microscopy (STEM) to investigate the local structure.  For imaging, we used the high angle annular dark field detector (HAADF), which yields a contrast nearly proportional to $Z^2$ ($Z$ = atomic number) and is therefore better suited for imaging vacancies and interstitials, and cations with different atomic numbers.  

\begin{figure}[ht]%
\includegraphics[width=0.9\columnwidth]{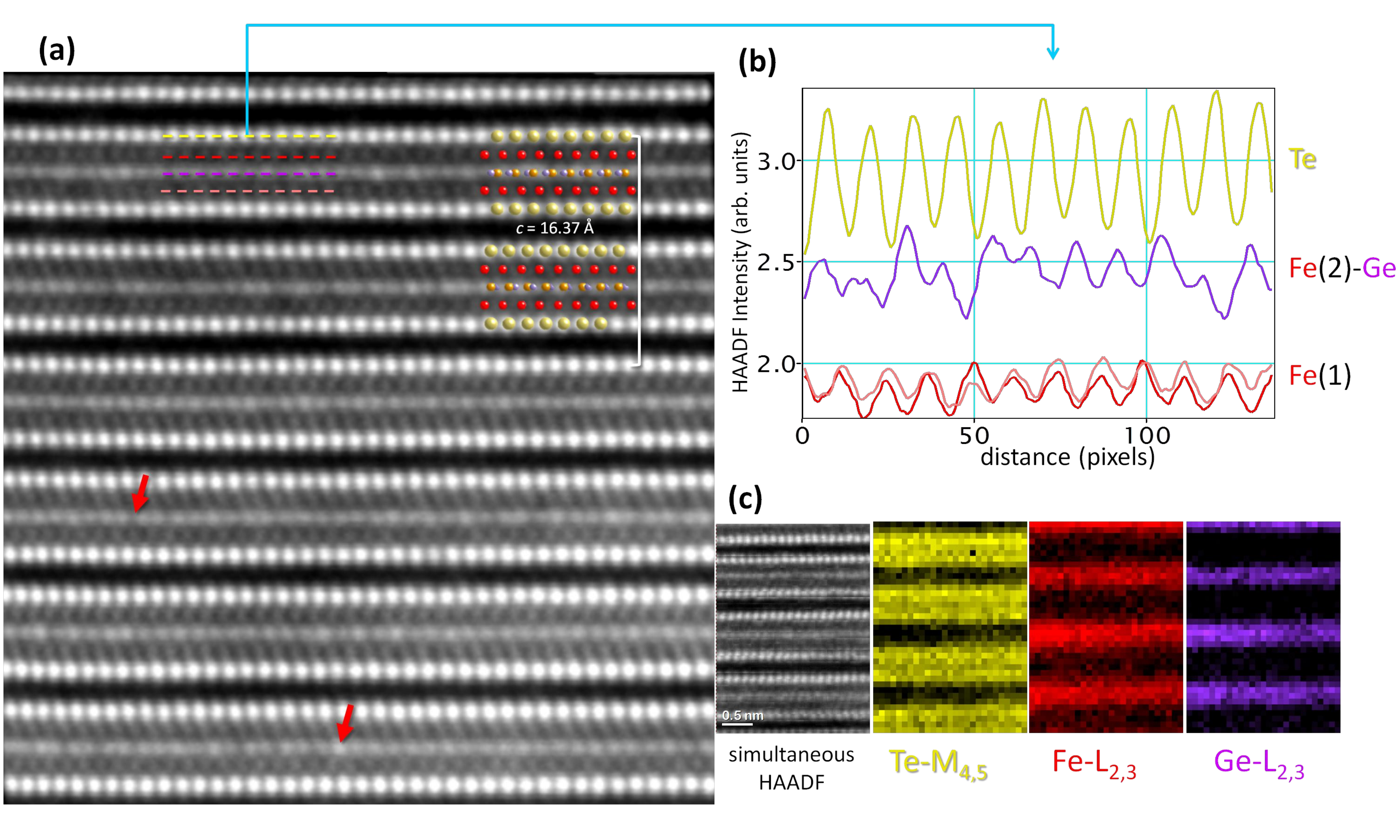} 
\caption{(color online) HAADF STEM image showing $Z$-contrast. The heaviest columns (Te) show up as the brightest spots and Fe columns show up with the dimmest contrast. Van der Waals gaps are clearly visible as black stripes. The inset shows an overlay of the lattice projection along [130]. (b) Profiles of the intensity along the dashed lines shown in (a) indicating significant disorder in the Fe(2)-Ge plane. (c) EELS integrated intensity maps showing the chemical signature of the alternating Te, Fe, and Fe-Ge planes.}%
\label{STEM}%
\end{figure}

A representative cross-sectional image with the electron beam parallel to [130] is shown in Fig.\,\ref{STEM}(a), and plan view images with the electron beam parallel to [001] were also collected but are not shown. In general, the crystals looked to be of high quality, lacking obvious two or three dimensional defects.  In the $Z$-contrast images, the van der Waals gaps between layers of Te atoms (brightest spots) can be easily identified and intensity is not observed between them, suggesting that interlayer Fe is not present in our crystals. In both cross-sectional and plan view images, we observe a variation in the intensity of the atomic columns containing Fe(2) and Ge, which is consistent with the presence of Fe vacancies.  The intensity profiles in Fig.\,\ref{STEM}(b) reveal these variations, with the Fe(2)-Ge columns displaying a clear disruption of the pattern expected for the case of fully occupied Fe(2) sites.  Due to the difference in $Z$ between Fe ($Z$=26) and Ge ($Z$=32), a fully occupied structure would display higher-intensity peaks adjacent to lower-intensity peaks in a high-low-high sequence when viewed down [130].  The red arrows in the image highlight Fe and Ge columns with markedly lower or higher intensity compared to the average. Based on the images, we cannot exclude the possibility of Ge-Fe antisite defects or Ge vacancies in the Fe(2)-Ge planes. Due to the low vacancy concentration and the more delocalized nature of the EELS signal as compared to the HAADF signal, vacancies are not resolved in the EELS compositional maps shown in Fig.\,\ref{STEM}(c), which show a uniform distribution for the integrated intensities of the Te-M$_{4,5}$, Fe-L$_{2,3}$ and Ge-L$_{2,3}$ edges.

The images were observed to change upon continued exposure to the electron beam.  A movement of atoms and vacancies could be observed, with atoms eventually occupying the interlayer region and vacancies appearing to occupy sites besides Fe(2).  In order to avoid beam-induced hopping and preserve observation of the intrinsic structure, images were collected rapidly on unirradiated regions by summing 20 frames, each one acquired within 1\,s. EEL spectrum images were acquired using a spacing of 0.87 \AA\, and a dwell time of 0.2 s.  Beam-induced structural effects were also noted for Ni$_3$GeTe$_2$ in Ref.\,\citenum{Deiseroth2006}. 

EELS was utilized to look for differences in the oxidation state of the two Fe sites.  The ratio of the Fe L$_{3,2}$ peaks gives an indication regarding the oxidation state, though this can vary with details of the bonding environment.  We find an L$_{3,2}$ ratio that varies slightly between the Fe(1) and Fe(2) positions, which may suggest the Fe atoms are in slightly different oxidations states or carry a different moment.  

\begin{figure}[h]%
\includegraphics[width=.95\linewidth]{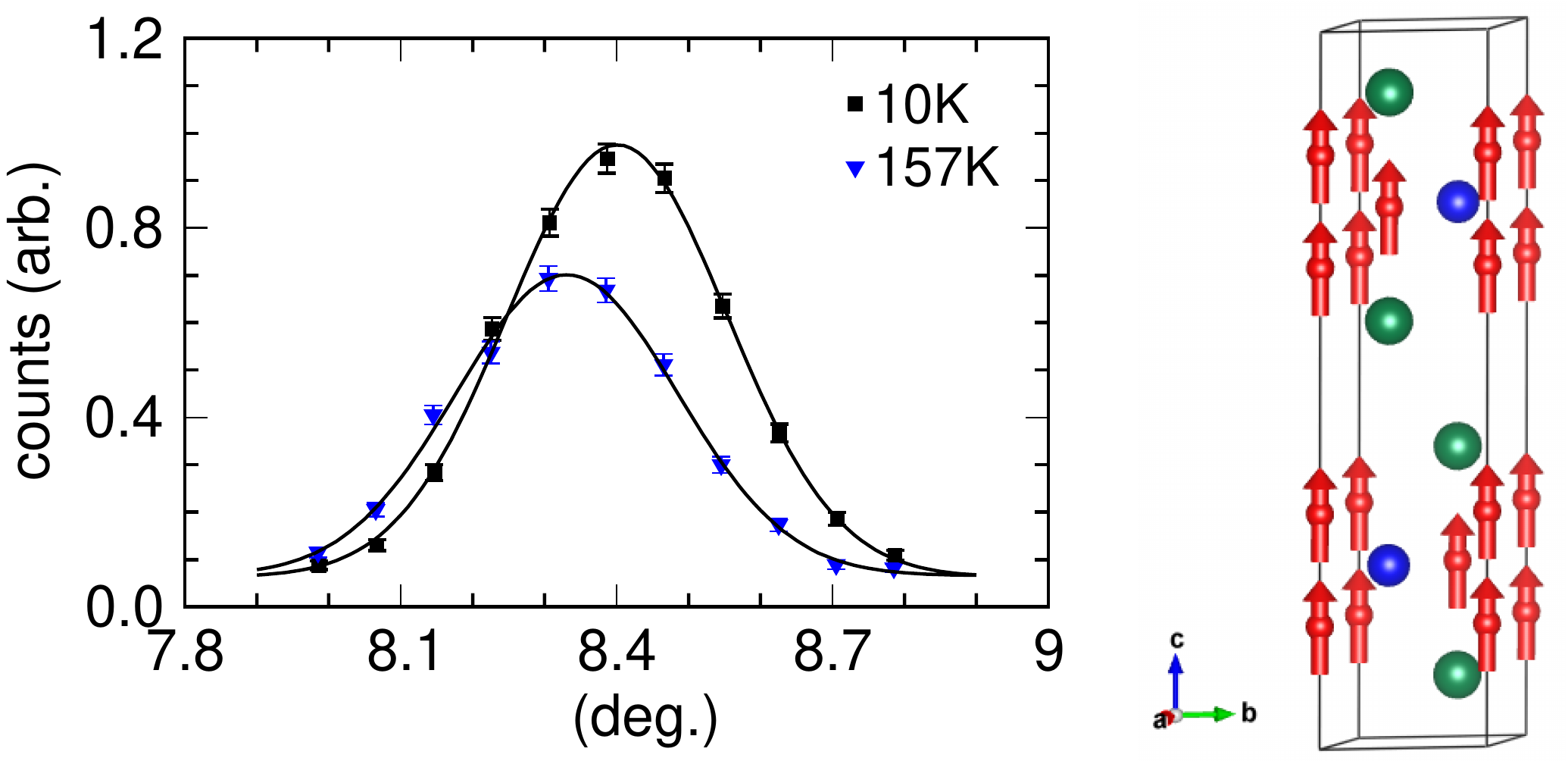}
\caption{(color online). Rocking curves of the 010 reflection from single crystal neutron diffraction on Fe$_{2.76}$Ge$_{0.94}$Te$_2$ at 10 K and 157 K. The solid lines are Gaussian fits to the peaks. The image on the right shows the magnetic structure obtained by refinement of neutron diffraction data collected at 4\,K (Fe = red, Te = green, Ge = blue).}%
\label{NDpeak}%
\end{figure}

The lack of evidence for occupation of Fe(3) in the STEM and diffraction studies may be due to the low-occupation of this position, which is likely to be further reduced with the overall decrease of Fe content in our crystals due to the flux growth.  Examination of Fe-rich crystals will offer the best chance for identifying interlayer Fe, if it exists, though even then the concentration will likely be very small and observation will be challenging.  Even at such low concentrations, though, the presence of interlayer Fe could strongly influence the physical properties.

\subsection{Neutron Diffraction}

Single crystal neutron diffraction was performed on a crystal obtained from the Fe$_2$GeTe$_4$ flux.  This flux composition was found to produce the largest crystals, and thus it was investigated in the most detail (including  magnetization and transport measurements below).  The crystal structure was refined using data collected at 220\,K, and the magnetic structure was obtained from data collected at 4\,K.  As discussed below, powder neutron diffraction measurements were also performed to verify the magnetic structure and investigate variations in the Fe moment(s) as a function of composition.

\begin{figure}[h]%
\includegraphics[width=0.9\columnwidth]{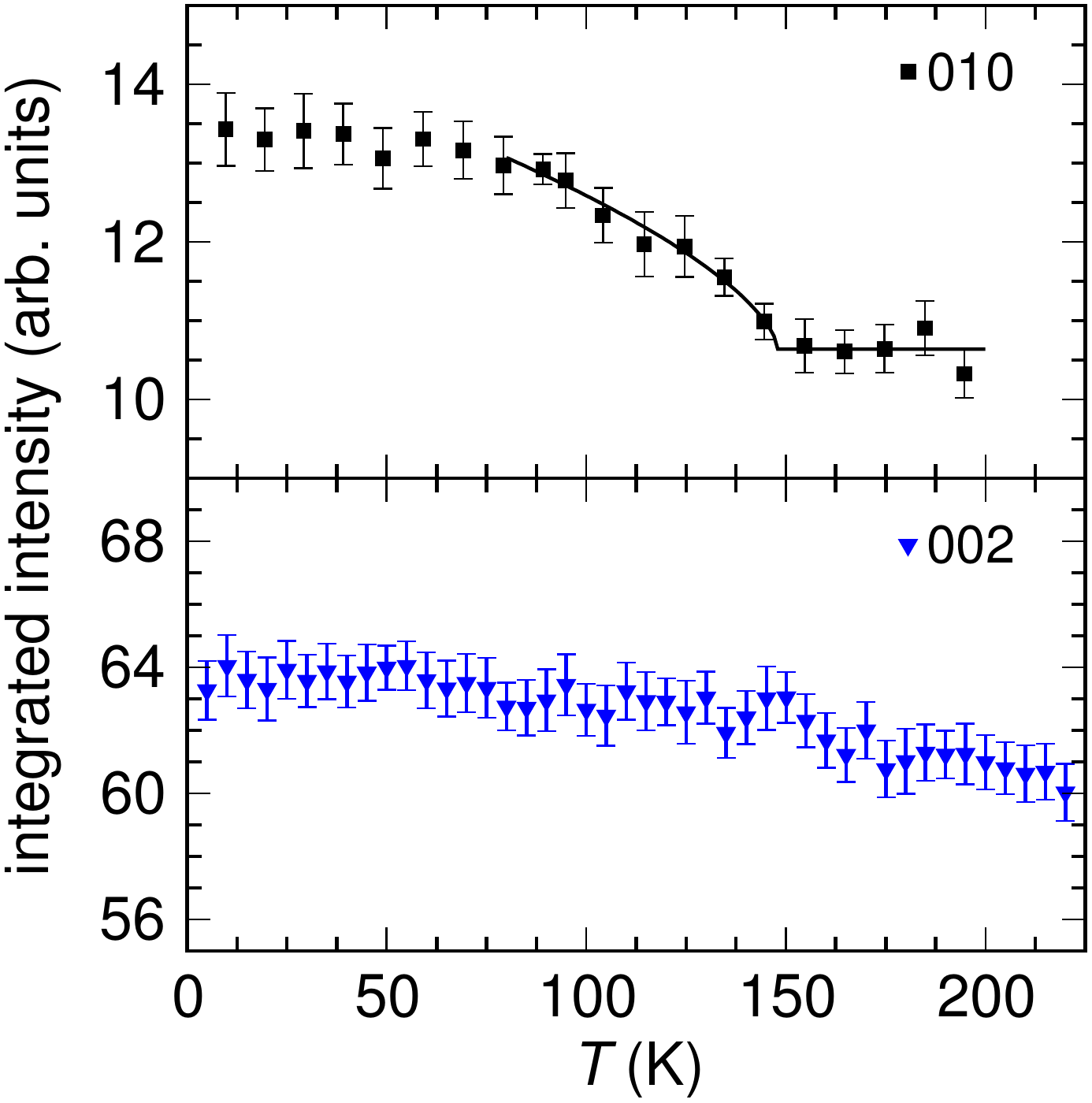}%
\caption{(color online). Integrated intensities of Bragg peaks from single crystal neutron diffraction on Fe$_{2.76}$Ge$_{0.94}$Te$_2$.  The increase in intensity for the 010 peak below $T_{\textrm{c}}$ originates from the magnetic order along $c$, whereas the intensity of the 002 peak does not change significantly with temperature.}%
\label{NDT}%
\end{figure}

Figure \ref{NDpeak} shows representative rocking curves obtained during the single crystal neutron diffraction data collection.  As shown, there is strong temperature dependence that indicates the presence of magnetic order below 157\,K. The nuclear refinement is consistent with the single crystal and powder x-ray diffraction data.  We refine vacancies at the Fe(2) and Ge sites, and do not observe strong evidence for Fe substituting for Ge.  Consistent with the above discussion, we do not detect any Fe between the Te-Te layers.  The refined composition is Fe$_{2.76(4)}$Ge$_{0.94(4)}$Te$_2$ with $R_F$=4.11 and $\chi^2$=0.423.  The site occupancies were fixed to these values for the lower temperature refinements. 

An ordered moment along $\textit{\textbf{c}}$ of 1.11(5)$\mu_B$/Fe is obtained from refinement of the single crystal data at 4\,K, and a schematic of the magnetic structure is shown in Fig.\,\ref{NDpeak}.  Our refinement does not indicate any significant difference in the moments on the two Fe positions for the Fe-deficient crystal.  If we allow both moments to refine separately, we obtain 1.07(11)$\mu_B$/Fe(1) and 0.9(5)$\mu_B$/Fe(2). We note that the errors on the moment values increase appreciably when both are allowed to refine separately, and this gives values consistent with that for fixed Fe(1) and Fe(2).  Thus, within the limits of this data, we have no reason to suspect that the different sites carry significantly different moments at this Fe concentration.  Very recently, based on powder neutron diffraction data from Fe$_{2.9}$GeTe$_2$, the ratio of moments between Fe(1) and Fe(2) was found to be 1.25 at 1.5\,K.\cite{Verchenko2015}   For comparison, in Fe$_{1.76}$Ge the ordered moments lie in the $ab$-plane, and an average moment of 1.56$\pm$0.2$\mu_B$/Fe was reported based on neutron diffraction.\cite{Katsuraki1964} There have been contradictory reports regarding a variation of the moments between the two Fe sites in Fe$_{2-x}$Ge.\cite{Katsuraki1964,Germagnoli1966,Albertini1998}  The more recent M\"{o}ssbauer results have suggested that a larger moment resides on the Fe(2) site, and perhaps similar experiments on Fe$_3$GeTe$_2$ may provide additional insight into the roles of the different Fe environments.  The difference in the easy axis between Fe$_3$GeTe$_2$ and Fe$_{1.67}$Ge is most likely caused by the increased chemical anisotropy associated with the inclusion of the Te-Te layer into the Fe$_3$GeTe$_2$ structure.

We confirmed the ferromagnetic ordering temperature and orientation of the moments by tracking the 010 and 002 Bragg peaks, as shown in Fig.\,\ref{NDT}.  The power law fit between 80\,K and 180\,K in Fig.\,\ref{NDT} yields $T_{\textrm{c}}$ = 148(3)\,K, which is consistent with the bulk magnetization measurements.  The intensity of the 010 peak increases when the moments lie perpendicular to the 010 scattering vector.  Therefore, the increase in intensity below $\approx$ 150\,K shown in Fig.\,\ref{NDT}(a) demonstrates that the moments do not lie along the $b$-axis, and when combined with the temperature-independent behavior of 002 intensity we verify that the moments lie along $\textit{\textbf{c}}$.  This shows that there is not any significant spin canting or reorientation as temperature decreases, which is also confirmed with the powder neutron diffraction measurements.  In addition, we did not observe any significant change in the nuclear structure across the magnetic transition.

\begin{figure}[h]%
\includegraphics[width=0.9\columnwidth]{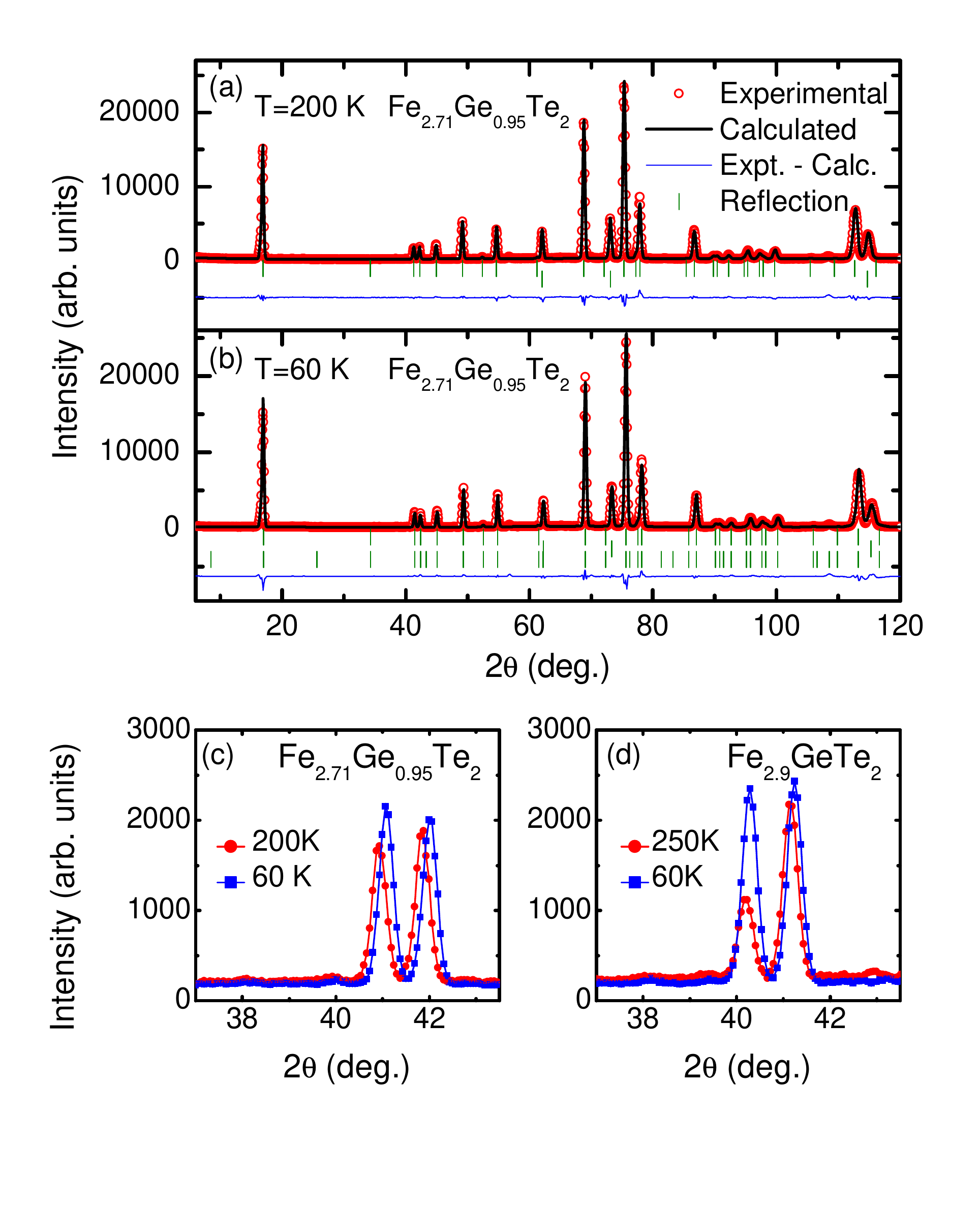}%
\caption{(color online). Neutron powder diffraction at (a) 200\,K and (b) 60\,K with refinements included and samples labeled by the refined compositions. The reflections from top to bottom correspond to the structural, aluminum can and magnetic reflections. The most intense magnetic signal was observed in the region shown in figures (c) and (d). Note the much increased magnetic intensity in the Fe$_{2.9}$GeTe$_2$ sample compared to the more Fe-deficient Fe$_{2.71}$GeTe$_2$.}%
\label{NPD}%
\end{figure}

\begin{table*}[ht!]
  \caption{Summary of refinements of powder neutron diffraction data. Refined compositions are the result of refining the occupation of the Fe(2) position.  Refined moments are aligned along the $c$-axis.}
  \label{NPDTable}
  \begin{tabular}{cccccccc}
    \hline
        Nominal Composition & $T$      & Refined Fe  & $a$        & $c$         & moment Fe(1) & moment Fe(2) & $R_P$\\
         										& (K)      &             & (\AA)      & (\AA)       & ($\mu_B$)    & ($\mu_B$)    &  \\
    \hline
        Fe$_{3}$GeTe$_2$    &  250\,K  & 2.904(8)    & 4.01749(8) & 16.33990(8) & -             & -           &  3.02     \\
        Fe$_{3}$GeTe$_2$    &  60\,K   & 2.904(8)    & 4.00938(3) & 16.2850(2)  & 2.18(10)      & 1.54(10)    &  3.41      \\
        Fe$_{2.75}$GeTe$_2$ &  200\,K  & 2.71(3)     & 3.95001(7) & 16.4019(4)  & -             & -           &  3.42       \\
        Fe$_{2.75}$GeTe$_2$ &  60\,K   & 2.71(3)     & 3.93628(4) & 16.3535(3)  & 1.4(1)        & 1.4(1)      &  5.25 \\
   \hline
  \end{tabular}
\end{table*}
During the single crystal neutron diffraction measurements, we scanned along H and L and found no additional non-integer (H,K,L) intensity through the magnetic transition of 150\,K.  Additionally, in the neutron powder diffraction data shown in Fig.\,\ref{NPD}, there is no evidence of scattering at non-integer (HKL) positions in going from above the Curie temperature $T_C$ to below $T_C$.   This strongly suggests that there is no deviation from the observed ferromagnetic structure for both Fe-rich and Fe-deficient samples.  We did notice some diffuse scattering beneath the Bragg peaks in our single crystal neutron diffraction, and this scattering was persistent at temperatures well above the magnetic ordering.  As such, this is likely related to crystalline defects and requires further investigation (note that stacking faults were not observed in our STEM).

Neutron powder diffraction was performed on polycrystalline samples of nominal compositions Fe$_{3}$GeTe$_2$ and Fe$_{2.75}$GeTe$_2$ (see Fig.\,\ref{NPD} where refined compositions are utilized).  In both cases, no deviation from the magnetic structure obtained from single crystal neutron diffraction was observed, indicting similar ferromagnetism regardless of Fe concentration.  The refinement results are summarized in Table\,\ref{NPDTable}.  The Fe-rich sample, with a refined composition of Fe$_{2.90}$GeTe$_2$, was found to have a stronger magnetic contribution to the Bragg peaks and the refinement yielded a larger ordered moment on Fe(1) than on Fe(2) (the raw data for this sample are shown in the Supplemental Materials).  Fixing the moments to be the same on the Fe(1) and Fe(2) sites did not provide suitable fits to the data for this sample.  This result is consistent with a recent publication containing neutron powder diffraction on a sample of Fe$_{2.90}$GeTe$_2$.\cite{Verchenko2015}  The behavior is different in the Fe-deficient sample, which has a refined composition of Fe$_{2.71}$Ge$_{0.95}$Te$_2$.  In this sample, the magnetic contribution to the diffraction peaks is smaller and appears similar for both Fe(1) and Fe(2).  The refined moment is slightly larger than that obtained on the single crystal (similar composition), but the trend for similar moments on the Fe(1) and Fe(2) sites is consistent.  This suggests that vacancies on the Fe(2) site suppress the magnetism on both Fe sites.  This could be related to the changes in structural parameters, though disorder and/or dilution effects may also be dominant.

\subsection{Magnetization}

\begin{figure}[h]%
\includegraphics[width=0.9\columnwidth]{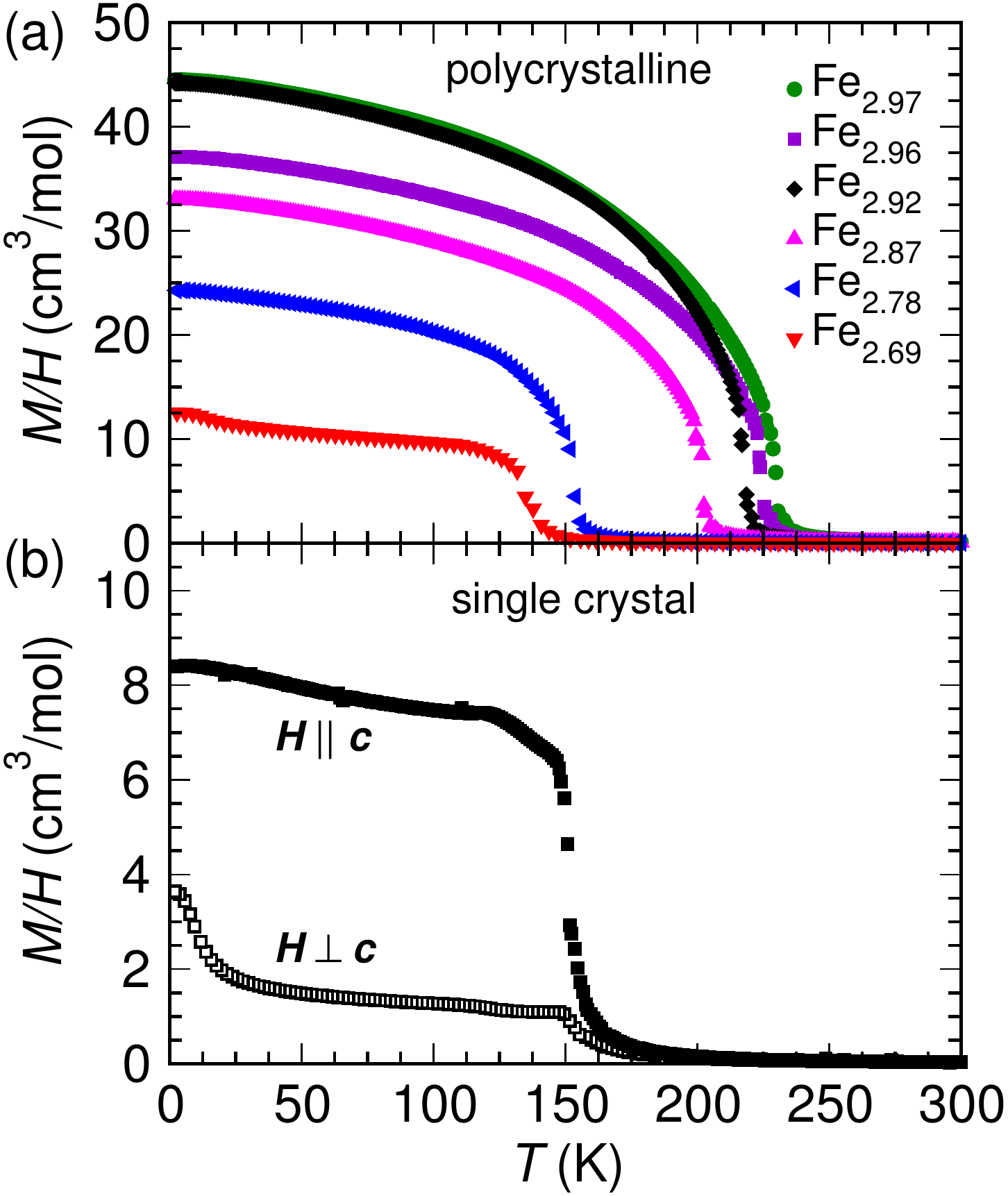}%
\caption{(color online) (a) Temperature dependence of the magnetization $M$ divided by applied field $H$ for polycrystalline Fe$_{3-x}$GeTe$_2$ materials with various Fe concentrations as indicated by the refined Fe content in the legend, and (b) anisotropic magnetization of flux-grown Fe$_{2.76}$Ge$_{0.94}$Te$_2$.  All data were collected upon cooling in an applied field of $H$=100\,Oe.}
\label{MTpoly}
\end{figure}

Magnetization measurements were performed on the polycrystalline materials to determine their Curie temperatures and correlate this with structure and composition.  Results of the temperature-dependent magnetization $M$ measurements are shown in Fig.\,\ref{MTpoly}(a), where refined Fe contents are used in the legend.  We have used Gaussian-CGS units for the magnetization results.

The data in Fig.\,\ref{MTpoly}(a) demonstrate that the Curie temperature decreases with decreasing Fe content.  The reduced $T_{\textrm{c}}$ with increasing vacancies on the Fe(2) site may be caused by a disruption of the magnetic exchange with increasing disorder and magnetic dilution via vacancies.  There may also be a structural component, as we observe that the reduced $T_{\textrm{c}}$ correlates with the expansion of $c$, an increase in Fe(1)-Fe(1) bond distance, and decrease in Fe(1)-Fe(2) bond distances.   A similar reduction in $T_{\textrm{c}}$ is observed for Fe$_{2-x}$Ge materials as $x$ increases, though, where a more typical decrease in the lattice parameters is observed with increasing $x$.\cite{Kanematsu1965}  Investigating the pressure-dependence of $T_{\textrm{c}}$ or the anisotropy of the magnetic excitation spectra may provide further insight, as would theoretical calculations into the dependence of $T_{\textrm{c}}$ on $c$.  

Anisotropic magnetization data for single crystalline Fe$_{3-x}$GeTe$_2$ are shown in Fig.\,\ref{MTpoly}(b).  These data demonstrate that the easy axis for magnetization is along the crystallographic $c$-axis, which is consistent with our neutron diffraction and Ref.\,\citenum{Chen2013}.  The shape of $M$($T$) evolves with decreasing Fe content.  In the Fe-rich samples, $M$($T$) increases smoothly while cooling.  In contrast, for the most Fe-deficient sample, $M$($T$) has a kink-containing shape near $T_C$ and essentially plateaus slightly below $T_C$.
The $M$($T$) data reported on the vapor transport crystals, with $T_C\approx220\,K$, has a temperature dependence similar to that shown in Fig.\,\ref{MTpoly}(b).  Therefore, the non-power law behavior observed near $T_C$ in the single crystals may be linked to domain wall formation and movement.  It is certainly possible that different growth conditions produce different Fe(2)/Fe(3) contents, which could also modify the properties independently.  We also confirmed that Ni$_{3-x}$GeTe$_2$ appears to be a Pauli paramagnet.

\begin{figure}[ht!]%
\includegraphics[width=0.9\columnwidth]{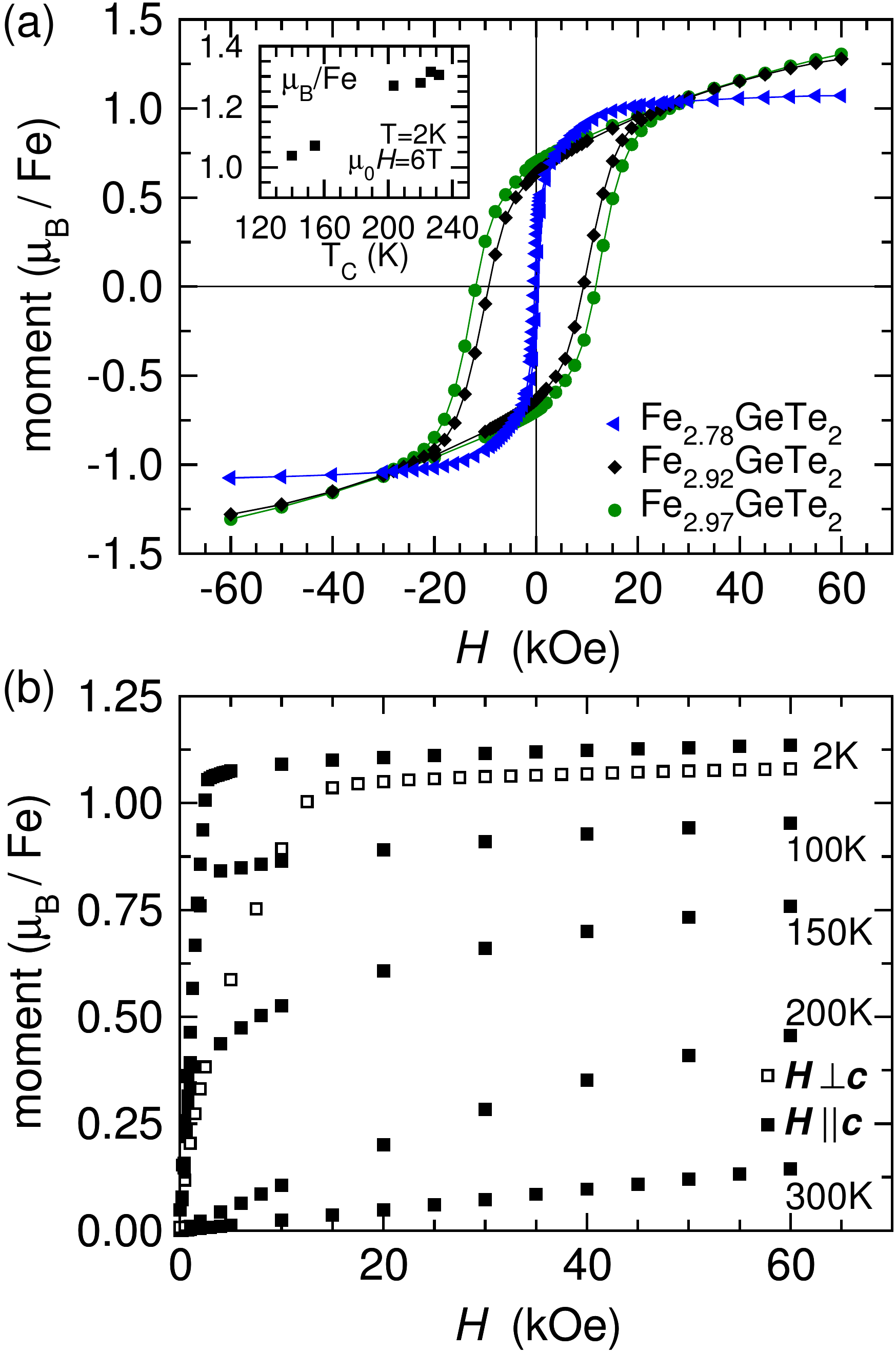}
\caption{(color online). (a) Magnetization loops at 2\,K for three compositions of polycrystalline Fe$_{3-x}$GeTe$_2$ (refined values provided), and the inset shows the magnetization at 2\,K and 60\,kOe for polycrystalline samples.  (b) Magnetization versus applied field for single crystalline Fe$_{2.76}$Ge$_{0.94}$Te$_2$ at various temperatures, with the anisotropy demonstrated at 2\,K.}%
\label{MH}%
\end{figure}

\begin{figure}[ht!]%
\includegraphics[width=0.9\columnwidth]{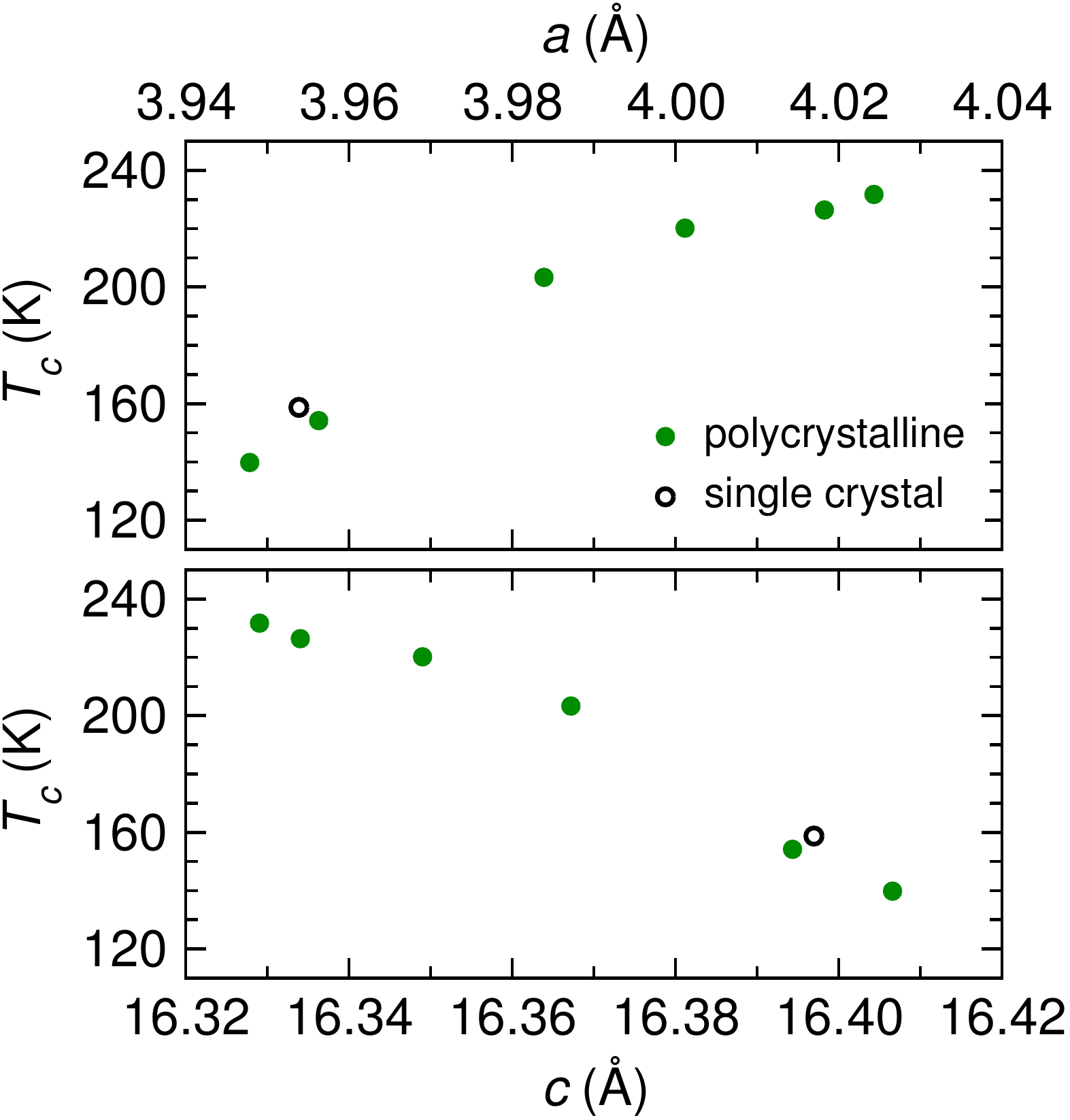}%
\caption{(color online). Curie temperature as a function of lattice parameters for Fe$_{3-x}$GeTe$_2$.}%
\label{Tc}%
\end{figure}

Isothermal magnetization data are shown in Fig.\,\ref{MH}.  In panel (a), data for three polycrystalline samples are shown while panel (b) contains data for an oriented single crystal.  The isothermal magnetization measurements for polycrystalline samples reveal some interesting trends with composition.  It is clear that the Fe-deficient samples have a much lower remanent magnetization and very little coercivity.  Also, the moment essentially saturates at high fields for the Fe-deficient samples while a linear rise with field is observed for the Fe-rich samples.  The linear increase at higher fields for Fe$_{2.97}$GeTe$_2$ may be related to an increased anisotropy field at higher Fe content.  Our low-Fe content crystals have an estimated anisotropy field of 15\,kOe (or 1.5\,T), as shown in Fig.\,\ref{MH}(b).  An anisotropy field of $\approx$5\,T was observed at 10\,K for crystals grown by vapor transport,\cite{Chen2013} and the reported $T_C\approx220$\,K suggests a large Fe content in those crystals.  The linear rise at large fields may be also be from some paramagnetic ions, such as an impurity or interlayer Fe.  We note that our Fe$_{2.76}$Ge$_{0.94}$Te$_2$ crystals did not reveal any unexpected behavior when measured to 120\,kOe.

The `saturation magnetization' of all polycrystalline Fe$_{3-x}$GeTe$_2$ samples is shown as an inset in Fig.\,\ref{MH}(a) and values are listed in the summary of samples provided as Table \ref{samples}.  The saturation magnetization is taken as the value of the magnetization (in $\mu_B$/Fe) obtained at 2\,K and 60\,kOe (refined compositions are used for unit conversion).  This does not represent a true saturation magnetization for all samples due to the linear increase in $M$ at large $H$.  We clearly see that higher $T_{\textrm{c}}$ (higher Fe content) correlates with larger induced moments for a given $T$ at large applied fields, consistent with our powder neutron diffraction results.  $T_{\textrm{c}}$ was defined using the intercept of the steepest tangent.  The effective moments calculated from the susceptibility ($\chi$) between 250 and 360\,K were found to vary between 3.9(2) and 4.9(1)$\mu_B$/Fe, with higher $T_{\textrm{c}}$ generally corresponding to larger effective moments.  We used a standard Curie-Weiss law ($\chi = C/(T-T_{CW}$)) and fit data collected on cooling in an applied field of 1\,kOe.  Curie-Weiss temperatures $T_{CW}$ obtained from these fits agreed well with the Curie temperatures obtained from measurements at lower fields.

\begin{table*}[ht!]
  \caption{Summary of samples and magnetic properties for polycrystalline Fe$_{3-x}$GeTe$_2$. Rietveld compositions originate in refinement of powder diffraction data while EDS values are for large area EDS analysis that may include impurities.}
  \label{samples}
  \begin{tabular}{ccccccccc}
    \hline
        \multicolumn{3}{c}{Fe composition} & $a$ & $c$  & $T_{\textrm{c}}$ & $\mu_{sat}$ & $\mu_{eff}$ & $T_{CW}$\\
        
        nominal & Rietveld  & EDS  & (\AA) & (\AA) & (K) & ($\mu_B$/Fe) & ($\mu_B$/Fe) & (K) \\
    \hline
        3.10   &   2.97(2)  & 3.07(2) & 4.0244(1) & 16.3293(5) & 232 & 1.31 & 4.9(1) & 225.6(2) \\
        3.00   &   2.96(1)  & 3.06(3) & 4.0184(1) & 16.3344(4) & 226 & 1.32 & 4.8(1) & 221.7(1) \\
        2.90   &   2.92(1)  & 2.92(3) & 4.0013(1) & 16.3494(4) & 220 & 1.28 & 4.4(1) & 218.9(1) \\
        2.85   &   2.87(1)  & 2.88(2) & 3.9840(1) & 16.3676(3) & 203 & 1.27 & 4.4(1) & 203.2(1) \\
        2.80   &   2.78(1)  & 2.81(2) & 3.9564(1) & 16.3947(4) & 154 & 1.08 & 4.5(1) & 143.5(3) \\
        2.75   &   2.69(4)  & 2.79(3) & 3.9475(4) & 16.405(1)  & 140 & 1.04 & 3.9(2) & 142.6(2) \\
   \hline
  \end{tabular}
\end{table*}

We calculated the Rhodes-Wohlfarth ratio (RWR) for our polycrystalline samples.  The RWR provides a quick means to characterize the degree to which a magnetic moment is localized.  RWR is defined as RWR = $p_{c}$/$p_{s}$, with $p_{c}$ obtained from the effective moment $p_{c}(p_{c}+2)=p_{eff}^2$.\cite{Wohlfarth1978,Moriya1979}  Physically, $p_{c}$ is the saturation moment expected from the effective moment calculated from the susceptibility in the paramagnetic phase (assuming Curie-Weiss behavior) and $p_s$ is the saturation moment obtained in the ordered state. RWR = 1 for localized systems and is larger in an itinerant system, with the ratio increasing as $T_{\textrm{c}}$ decreases for itinerant systems.  Here, we take $p_s$ as the magnetization obtained at 2\,K and 60\,kOe and calculate RWR values between 2.7 and 3.4 for our polycrystalline samples.  These values are fairly similar to the RWR = 3.8 reported in Ref.\,\citenum{Chen2013}.

The observation of RWR $>$ 1 in compounds with a low Curie temperature ($T_C \alt 500$\,K) suggests itinerant ferromagnetism is likely present.  In comparison to the values tabulated by Wohlfarth in 1978 and Moriya in 1979, these Fe$_{3-x}$GeTe$_2$ compounds lie in the region between localized and itinerant ferromagnetism.\cite{Wohlfarth1978,Moriya1979}  We do not observe a strong magnetoelastic effect at the transition, which is expected for large RWR (itinerant) systems.\cite{Wohlfarth1978} While we do not observe a strong increase in the RWR values as $T_{\textrm{c}}$ decreases, this can likely be attributed to the influence of vacancies on the structure and magnetism (multiple effects influencing both $T_{\textrm{c}}$ and RWR). Future measurements of the spin-waves via inelastic neutron scattering will aid in addressing the itinerant nature of this system, as would theoretical or experimental studies into the influence of pressure on $T_{\textrm{c}}$.

We have summarized the magnetization data as a plot correlating the lattice parameters with the Curie temperature, shown in Fig.\,\ref{Tc}.  In addition, Table \ref{samples} provides a summary of sample compositions determined using various methods as well as the magnetic properties. $T_{\textrm{c}}$ clearly decreases as Fe vacancies are introduced and the lattice responds with a decrease in the in-plane lattice parameter and a slight expansion along $\textit{\textbf{c}}$.  These results can be used as a guide to predict the composition needed to obtain a particular $T_{\textrm{c}}$ or as a means to expedite characterization, and must be considered when performing measurements on ultra-thin samples.  Figure\,\ref{Tc} shows that the flux-grown crystals do not behave unexpectedly based on the behavior of polycrystalline materials.

\subsection{Hall Effect and Seebeck Coefficients}

\begin{figure}
\includegraphics[width=0.9\columnwidth]{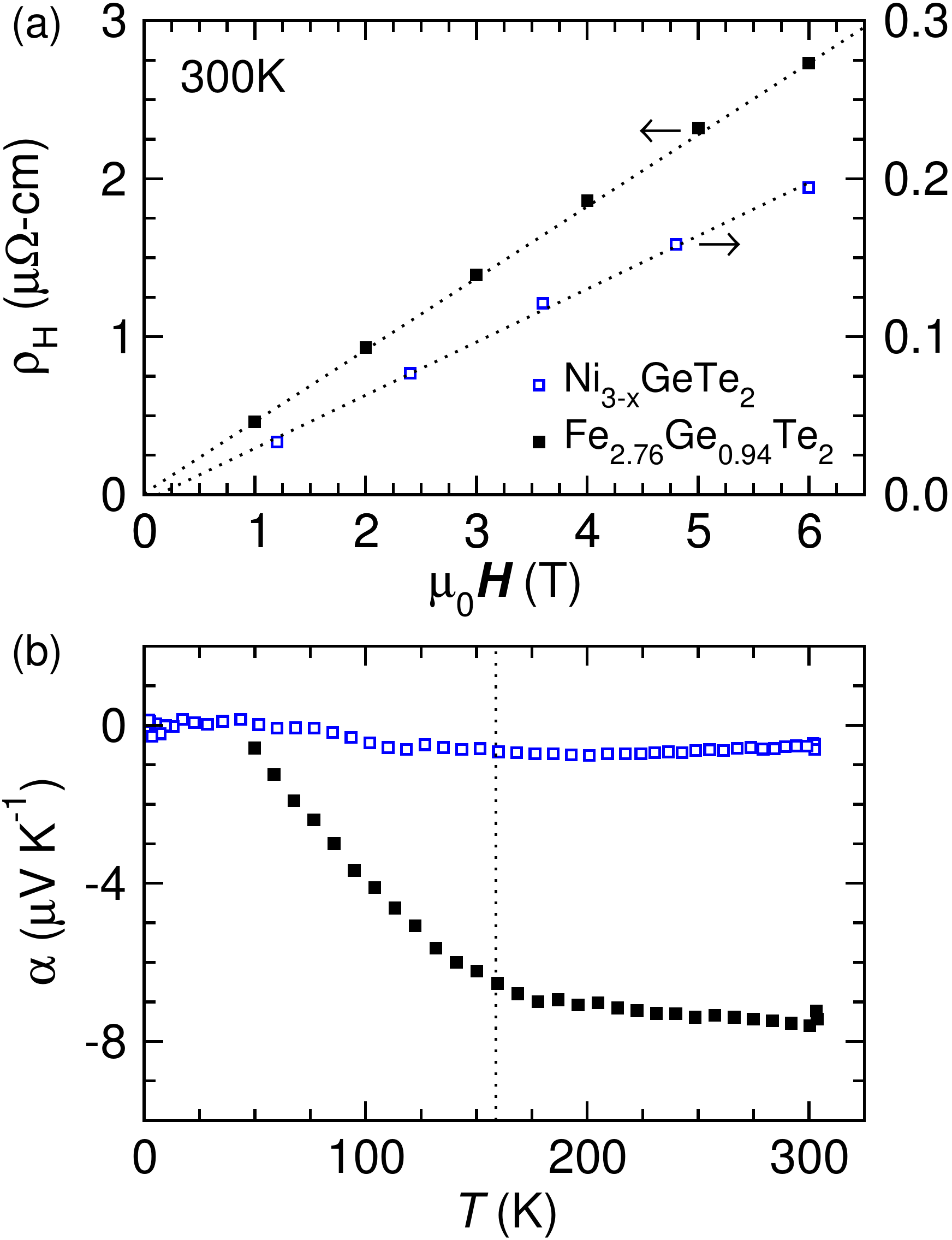}%
\caption{(color online) (a) Hall resistance as a function of applied field suggests $p$-type conduction in Fe$_{2.76}$Ge$_{0.94}$Te$_2$ and Ni$_{3-x}$GeTe$_2$ crystals, while in (b) the negative Seebeck coefficients indicates $n$-type conduction.  The dashed line in panel (b) indicates the Curie temperature of the Fe$_{2.76}$Ge$_{0.94}$Te$_2$ crystal.}%
\label{HallTTO}%
\end{figure}

We have utilized Hall effect and thermoelectric measurements to characterize the in-plane electrical transport in Fe$_{2.76}$Ge$_{0.94}$Te$_2$ and Ni$_{3-x}$GeTe$_2$ crystals.  These results, summarized in Fig.\,\ref{HallTTO}, reveal that both systems likely have multiple carrier types contributing to conduction.  We observe positive Hall coefficients and negative Seebeck coefficients, which would independently suggest $p$-type and $n$-type conduction, respectively.  The linear dependence of the Hall resistance on applied field prohibits a detailed analysis aimed at determining the contribution of each band/carrier-type.  Theoretical calculations would provide additional insight into the origin of this apparent multi-carrier transport.  We note that our flux-grown crystals of Ni$_{3-x}$GeTe$_2$ likely have $x$$\approx$0.6 based on EDS measurements.

At room temperature, the Hall coefficient $R_{\mathrm{H}}$ of Ni$_{3-x}$GeTe$_2$ is about an order of magnitude smaller than that of Fe$_{2.76}$Ge$_{0.94}$Te$_2$, which translates to a larger hole concentration in Ni$_{3-x}$GeTe$_2$ if a single-carrier model is used.  Specifically, at 300\,K the Hall carrier density $n_{\mathrm{H}}=1/R_{\mathrm{H}}e$ is approximately 1.8$\times$10$^{22}$cm$^{-3}$ for Ni$_{3-x}$GeTe$_2$ and approximately 1.9$\times$10$^{21}$cm$^{-3}$ for Fe$_{2.76}$Ge$_{0.94}$Te$_2$.  If both holes and electrons are present, as suggested by these results, the Hall coefficients could be artificially reduced and the carrier concentrations reported would be upper-limits to the actual number of holes in the system. A more complete compensation of charge carriers may be responsible for the smaller Hall coefficient of Ni$_{3-x}$GeTe$_2$, though we found Ni$_{3-x}$GeTe$_2$ to have about an order of magnitude lower electrical resistivity (Supplemental Materials).  As discussed in the Supplemental Materials, the Hall coefficient of Fe$_{2.76}$Ge$_{0.94}$Te$_2$ is strongly influenced by an anomalous Hall contribution below $\approx$200\,K.  Thermal conductivity and specific heat data for Fe$_{2.76}$Ge$_{0.94}$Te$_2$ and Ni$_{3-x}$GeTe$_2$ are also presented in the Supplemental Materials.\footnote{See Supplemental Material at [URL will be inserted by publisher] for powder neutron diffraction data, anomalous Hall effect data, electrical resistivity, thermal transport, and specific heat data.}

\section{Summary}

The availability of Fe$_{3-x}$GeTe$_2$ and other recently-developed, cleavable ferromagnets provides a starting point for the development of magnetically-active van der Waals heterostructures.  While such architectures will likely be designed to investigate specific physics or functionality, fundamental investigations will almost certainly yield unique spin structures or magnetotransport.  This work has demonstrated that the itinerant ferromagnetism in Fe$_{3-x}$GeTe$_2$ can be tuned by controlling the Fe content.   All manifestations of the magnetic interactions, from the Curie temperature to the local moment, are reduced as Fe vacancies are created and the in-plane lattice parameter decreases.  By mapping the magnetic phase diagram of Fe$_{3-x}$GeTe$_2$, this work has provided a foundation for future studies examining the influence of dimensionality on the magnetism in these van der Waals bonded materials.  While we have shown that the magnetic behavior can be controlled through total Fe content, chemical substitutions or intercalation may provide additional control over the magnetism and physical properties of this layered material.   Future work in this area will need to examine the stability of ultra-thin Fe$_{3-x}$GeTe$_2$ as a function of Fe content.  In addition, experiments under pressure and theoretical calculations will likely provide valuable information regarding the interactions between the electronic and magnetic structures of these multi-carrier, itinerant ferromagnets.

\section{Acknowledgements}

This work was supported by the U. S. Department of Energy, Office of Science, Basic Energy Sciences, Materials Sciences and Engineering Division.  Research performed at the High Flux Isotope Reactor at Oak Ridge National Lab was supported by the Department of Energy, Scientific User Facility Program.  We thank Radu Custelcean for assistance with single crystal x-ray diffraction data collection.


%

\renewcommand{\theequation}{S\arabic{equation}}
\renewcommand{\thefigure}{S\arabic{figure}}
\renewcommand{\thetable}{S\arabic{table}}
\renewcommand{\bibnumfmt}[1]{[S#1]}
\renewcommand{\citenumfont}[1]{S#1}

\setcounter{figure}{0}

\subsection{Supplemental Materials}

\begin{figure}[h]%
\includegraphics[width=0.9\columnwidth]{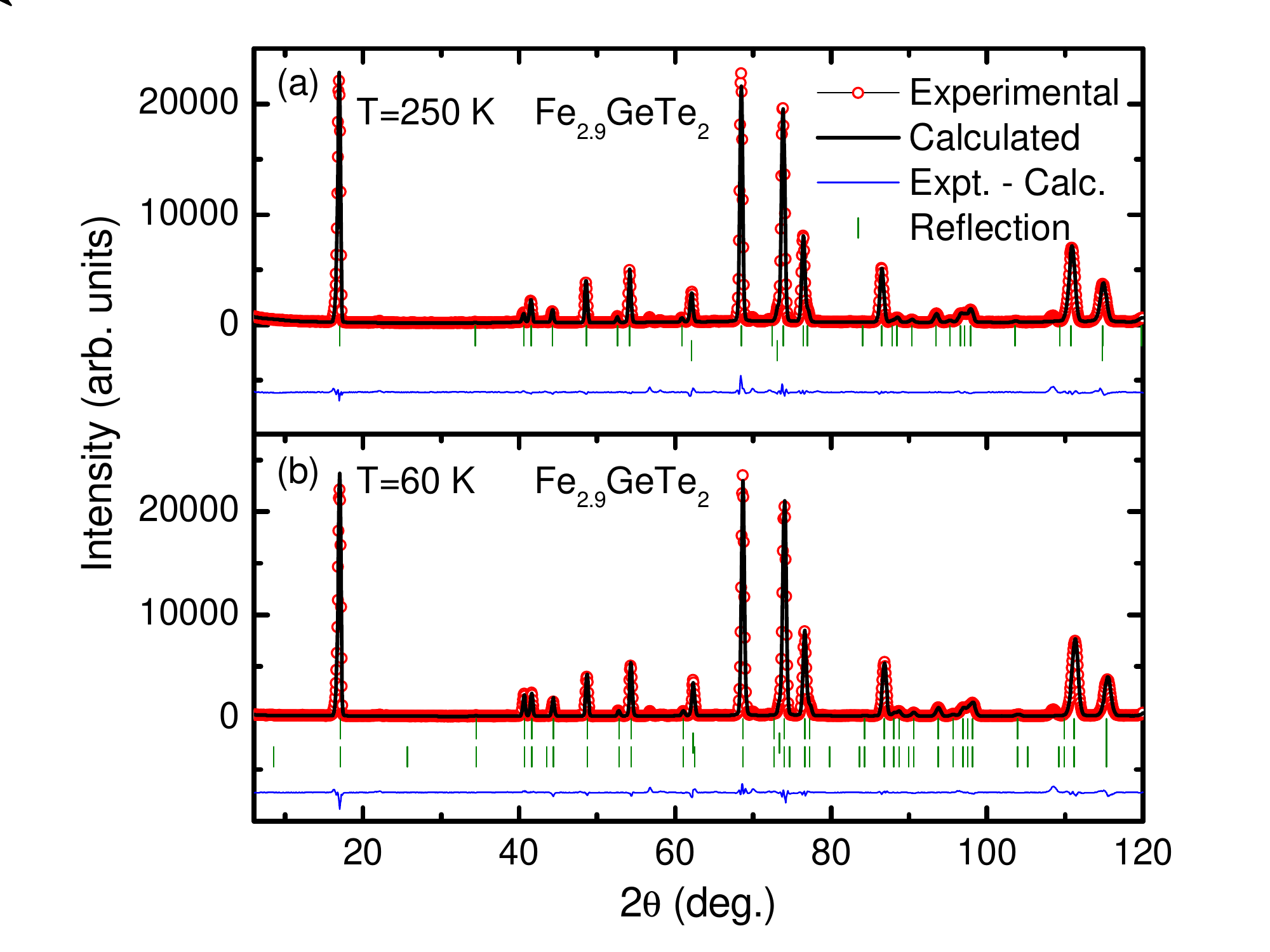}%
\caption{(color online) Neutron powder diffraction at (a) 250\,K and (b) 60\,K with refinements included and sample labeled by the refined composition. The reflections from top to bottom correspond to the structural, Aluminum can and magnetic reflections.}%
\label{NPD_SM}%
\end{figure}

Neutron powder diffraction data collected on a sample of nominal composition Fe$_3$GeTe$_2$ are shown in Fig.\,\ref{NPD_SM}.  For this sample, the refined composition was Fe$_{2.904(8)}$GeTe$_2$.  Refinement results are summarized in the main text.

\begin{figure}[h]%
\includegraphics[width=0.9\columnwidth]{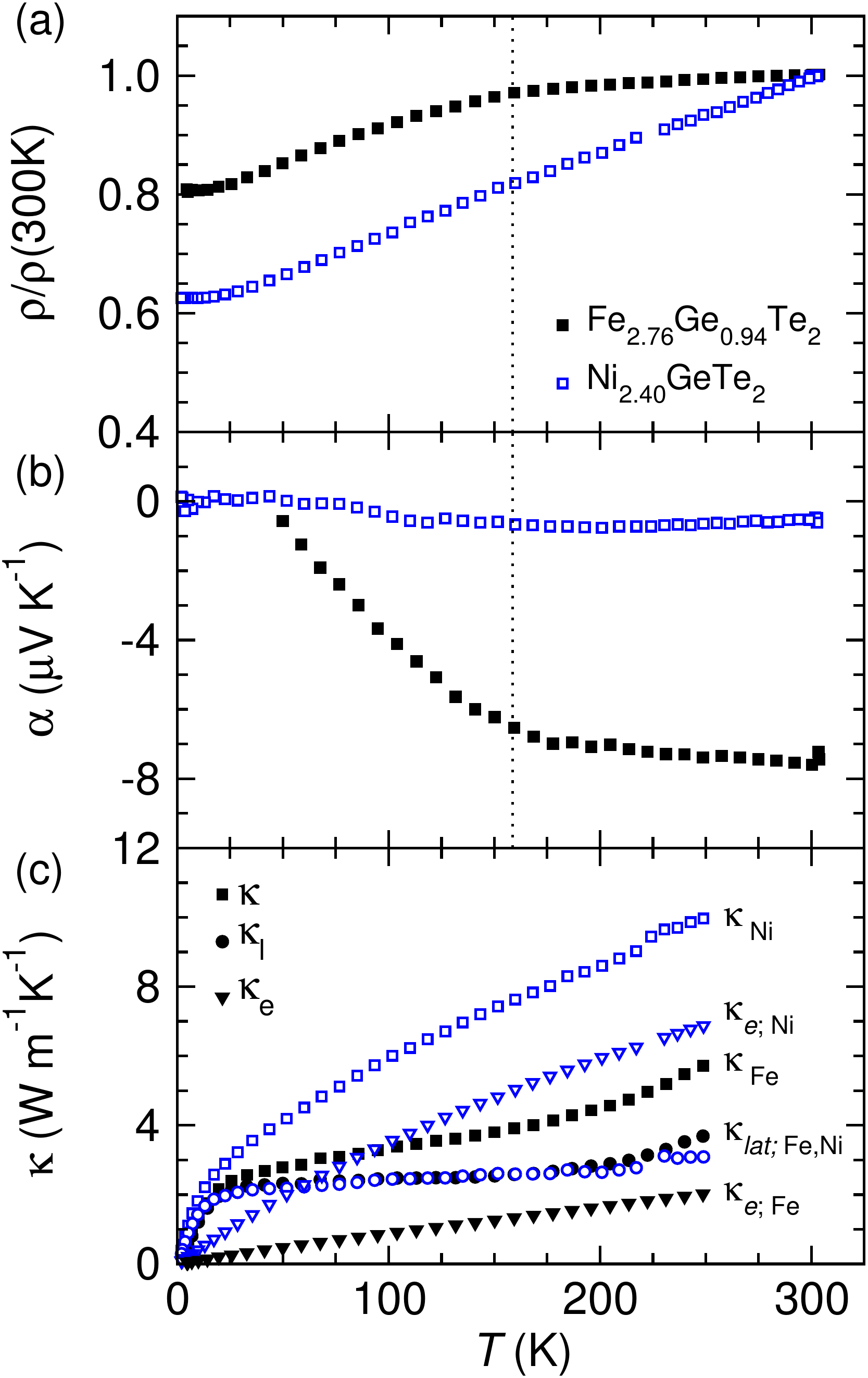}%
\caption{(color online) In-plane electrical and thermal transport for Fe$_{2.76}$Ge$_{0.94}$Te$_2$ and Ni$_{2.40}$GeTe$_2$ crystals. (a) Normalized electrical resistivity, (b) Seebeck coefficient and (c) thermal conductivity with estimates for the lattice and electronic contributions shown.  The closed black symbols are for Fe$_{2.76}$Ge$_{0.94}$Te$_2$ ($\rho$(300\,K)=3.0$\times$10$^{-4}\Omega$-cm) and the open blue symbols represent Ni$_{2.40}$GeTe$_2$ ($\rho$(300\,K)=9.50$\times$10$^{-5}\Omega$-cm).}%
\label{TTO}%
\end{figure}

With the production of large single crystals, we were able to perform in-plane thermal and thermoelectric transport measurements. Figure \ref{TTO} presents the electrical resistivity, Seebeck coefficient, and thermal conductivity of our flux-grown Fe$_{2.76(4)}$Ge$_{0.94(4)}$Te$_2$ crystals (in-plane transport).  This composition is obtained from refinement of single crystal neutron diffraction data. To facilitate a comparison, we have used the composition Ni$_{2.40(4)}$Ge$_{1.01(3)}$Te$_{2.00(3)}$ obtained from EDS. The electrical resistivity is about an order of magnitude lower for Ni$_{2.40}$GeTe$_2$ than for Fe$_{2.76}$Ge$_{0.94}$Te$_2$ (see caption of Fig.\,\ref{TTO}).  The Ni$_{2.40}$GeTe$_2$ sample also has a slightly lower residual resistivity ratio (RRR), as demonstrated by the plot of $\rho(T)/\rho$(300\,K) in Fig.\,\ref{TTO}(a). Transition metal vacancies and/or associated displacements of Ge likely provide significant charge carrier scattering that leads to small RRR in both of these systems.  

The Seebeck coefficient ($\alpha$) is small and negative for both samples.  The negative value implies electrons dominate conduction, and the smaller value for Ni$_{2.40}$GeTe$_2$ would imply a higher concentration of charge carriers (consistent with lower $\rho$) or a more complete compensation of electrons/holes.  Based on the Hall data discussed below, these appear to be multi-carrier metals.

The estimated lattice thermal conductivity $\kappa_{lat}$ is similar for Fe$_{2.76}$Ge$_{0.94}$Te$_2$ and Ni$_{2.40}$GeTe$_2$, as shown in Fig.\,\ref{TTO}(c).  These $\kappa_{lat}$ values were obtained using the Wiedemann-Franz law to estimate an electronic contribution $\kappa_e$ to the total thermal conductivity $\kappa$; the degenerate limit of the Lorenz number was assumed.  The values of $\kappa_{lat}$ are similar across the entire temperature range investigated and a low $T$ maximum is not observed.  The temperature dependence suggests vacancies likely dominate phonon scattering rates, and despite apparently different transition metal contents (based on our EDS results) the net result is a similar $\kappa_{lat}$. It is also possible that phonons are scattered by charge carriers at low $T$. 

The electrical properties respond to the ferromagnetic ordering in Fe$_{2.76}$Ge$_{0.94}$Te$_2$, as observed in Fig.\,\ref{TTO}(a,b) where the Curie temperature is indicated by the dashed line.  Below $T_C$, the electrical resistivity and Seebeck coefficient also begins to decrease more rapidly.  A decrease in $\rho$ below $T_C$ is commonly understood as a reduction in spin disorder scattering when the moments order.  The source for a decrease in the Seebeck coefficient is less clear.  The Seebeck coefficient is influenced by the scattering mechanisms as well as the carrier concentration and shape of the Fermi surface, the latter of which may be influenced by the magnetic ordering.  The magnetic scattering may asymmetrically influence the contributions of the various charge carriers, thereby influencing the Seebeck coefficient.  We have not noticed a strong response of the lattice to the magnetic ordering in Fe$_{2.76}$Ge$_{0.94}$Te$_2$.

\begin{figure}
\includegraphics[width=0.9\columnwidth]{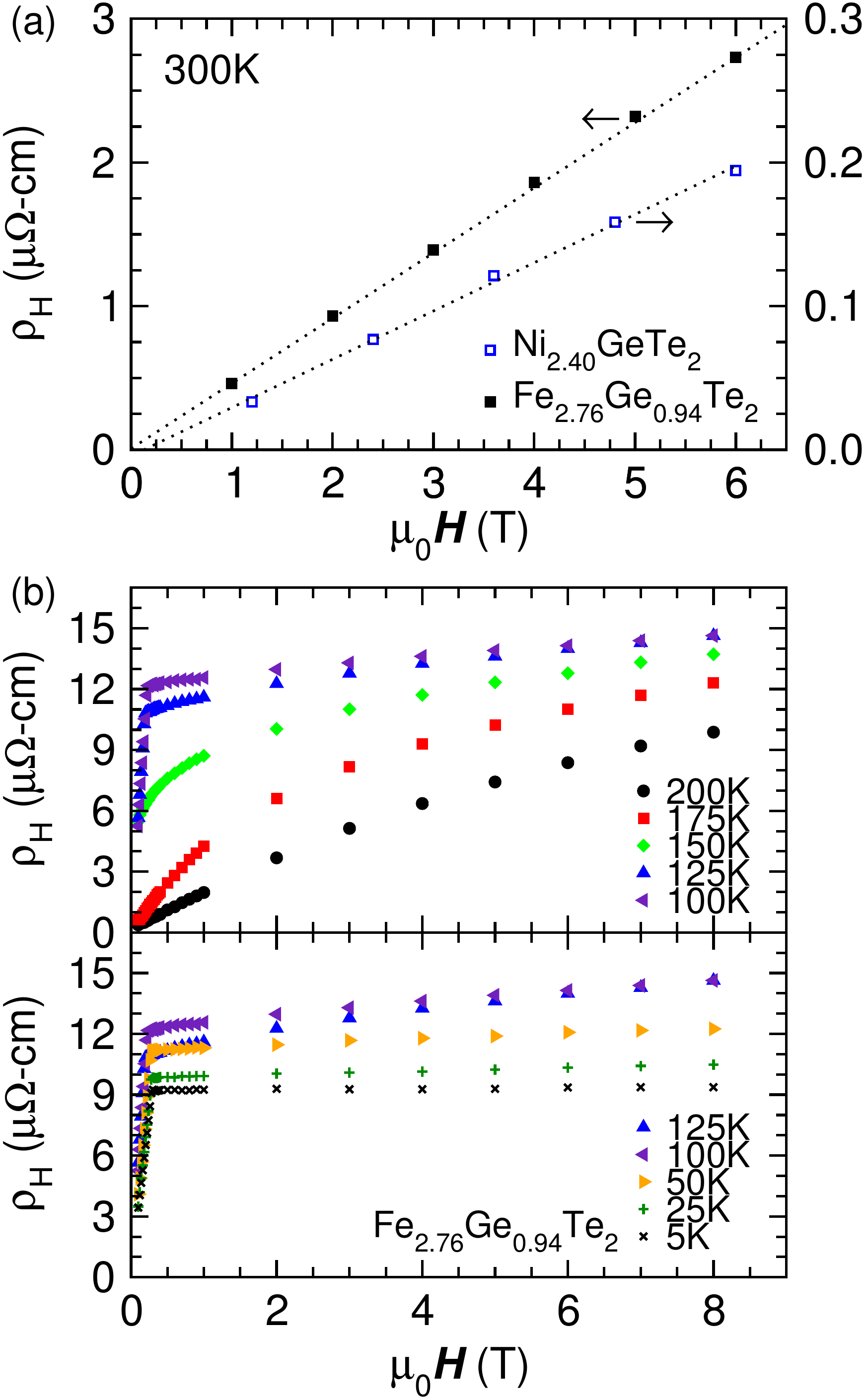}%
\caption{Hall effect data at (a) 300\,K for Fe$_{2.76}$Ge$_{0.94}$Te$_2$ and Ni$_{2.40}$GeTe$_2$ and in (b,c) the influence of an anomalous component brought about by the ferromagnetic ordering in Fe$_{2.76}$Ge$_{0.94}$Te$_2$ is demonstrated.}%
\label{Hall}%
\end{figure}

We performed Hall effect measurements to further characterize the electrical behavior of these systems.  The Hall coefficient is positive for both Fe$_{2.76}$Ge$_{0.94}$Te$_2$ and Ni$_{2.40}$GeTe$_2$, and the Hall resistance $\rho_{\small{\mathrm{H}}}$ is linear with magnetic fields up to at least 8\,T at 300\,K (Fig.\,\ref{Hall}(a)).  At room temperature, the Hall coefficient $R_{\mathrm{H}}$ of Ni$_{2.40}$GeTe$_2$ is about an order of magnitude smaller than that of Fe$_{2.76}$Ge$_{0.94}$Te$_2$, which translates to a larger carrier concentration in Ni$_{2.40}$GeTe$_2$ if a single-carrier model is used.  Specifically, at 300\,K the Hall carrier density $n_{\mathrm{H}}=1/R_{\mathrm{H}}e$ is $\approx$1.8$\times$10$^{22}$cm$^{-3}$ for Ni$_{2.40}$GeTe$_2$ and $\approx$1.4$\times$10$^{22}$cm$^{-3}$ for Fe$_{2.76}$Ge$_{0.94}$Te$_2$ using the linear fits shown in Figure\,\ref{Hall}(a).

The positive Hall coefficient suggests the dominant carriers are holes while the negative sign of the Seebeck coefficient suggests the dominant charge carriers are electrons. A detailed analysis of the contributions of each band is prohibited, however, due to the linearity of the Hall resistance with magnetic field. If both holes and electrons are present, as suggested by these results, the Hall coefficients could be artificially reduced and the carrier concentrations reported would be upper-limits to the actual number of holes in the system.   Ni$_{2.40}$GeTe$_2$ is more conductive, despite apparently having more defects, and thus an increase in the absolute number of charge carriers relative to Fe$_{2.76}$Ge$_{0.94}$Te$_2$ is likely.

The Hall effect of Fe$_{2.76}$Ge$_{0.94}$Te$_2$ is strongly influenced by the anomalous Hall contribution, as shown in Fig.\,\ref{Hall}(b,c).  For our crystals, with a Curie temperature of $\approx$ 150\,K, an influence of the anomalous Hall effect is observed below approximately 200\,K.  This is due to the strong polarization of the Fe moments with increasing field and decreasing temperature.  The Hall data follow the field dependence of the magnetization, which demonstrates that the non-linearity of $\rho_H$ is not due to multiband effects.  The regular and anomalous Hall coefficients have the same sign, and the current data are insufficient to analyze in detail due to the small contribution from the regular Hall coefficient as well as the non-saturating magnetization at high fields.  Qualitatively different results were obtained for the regular Hall coefficient when a detailed analysis was performed on data collected for different crystals, which had similar room temperature Hall coefficients (likely due to minor variations in magnetization between the crystals).

\begin{figure}
\includegraphics[width=0.9\columnwidth]{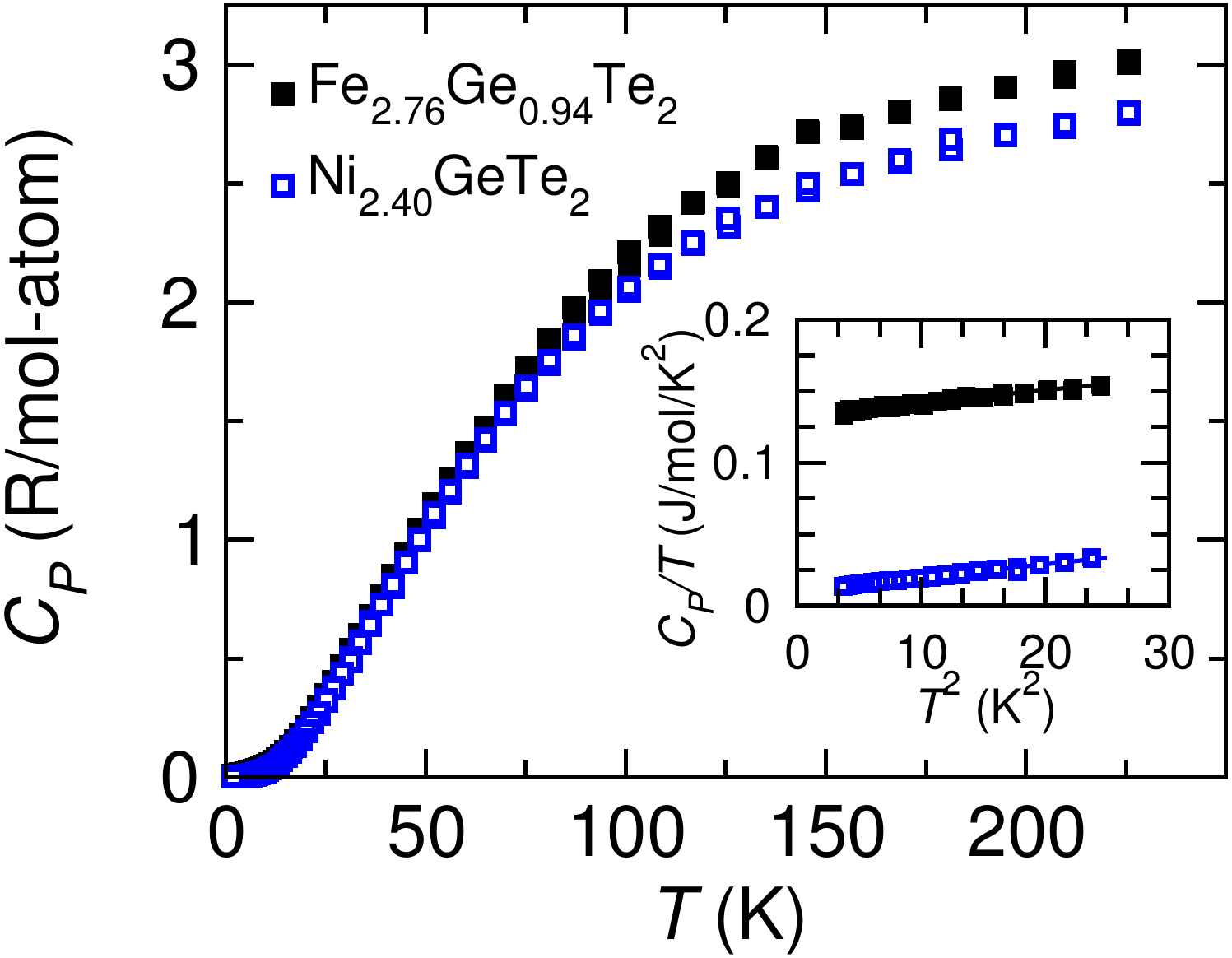}
\caption{(color online). Specific heat capacity of Fe$_{2.76}$Ge$_{0.94}$Te$_2$ and Ni$_{2.40}$GeTe$_2$, with inset showing low $T$ behavior and fits (thin solid lines).  A small anomaly is observed in Fe$_{2.76}$Ge$_{0.94}$Te$_2$ near $T_C$$\approx$150\,K.}%
\label{Cp}%
\end{figure}

Specific heat measurements were performed on single crystals of Fe$_{2.76}$Ge$_{0.94}$Te$_2$ and Ni$_{2.40}$GeTe$_2$, and the results are shown in Fig.\,\ref{Cp}.  A small anomaly is present in the region of the ferromagnetic transition of Fe$_{2.76}$Ge$_{0.94}$Te$_2$ while data for Ni$_{2.40}$GeTe$_2$ are smooth across the entire temperature range.   The measured values are approaching the high temperature limit of 3$k_B$/atom at 220\,K for both materials.  This is consistent with the Debye temperatures obtained from the low $T$ data, which were $\Theta_D$=224\,K and 234\,K for Ni$_{2.40}$GeTe$_2$ and Fe$_{2.76}$Ge$_{0.94}$Te$_2$, respectively.   The slightly smaller $\Theta_D$ for Ni$_{2.40}$GeTe$_2$ could be due to increased softening associated with a higher carrier density or a higher vacancy concentration, though potential error associated with the sample compositions precludes such a conclusion.  At low temperatures, the Debye temperature $\Theta_D$ is obtained from a plot of $C_P/T$ versus $T^2$ plot where the slope is $\frac{12\pi^4 R N_{at}}{5 \Theta_D^3}$ and $N_{at}$ is the number of atoms per formula unit.  

The electronic coefficient to the specific heat is significantly larger for Fe$_{2.76}$Ge$_{0.94}$Te$_2$ than for Ni$_{2.40}$GeTe$_2$.  We obtain a Sommerfeld coefficient $\gamma$ = 132.8\,mJ/mol/K$^2$ for Fe$_{2.76}$Ge$_{0.94}$Te$_2$ and 10.2\,mJ/mol/K$^2$ for  Ni$_{2.40}$GeTe$_2$.  The large $\gamma$ for Fe$_{2.76}$Ge$_{0.94}$Te$_2$ is either due to a mass enhancement from correlations or due to a contribution from spin fluctuations.  We note that our $\gamma$ is similar to that reported for vapor transport grown Fe$_3$GeTe$_2$ crystals.\cite{Chen2013}


\begin{thebibliography}{32}%
\makeatletter
\providecommand \@ifxundefined [1]{%
 \@ifx{#1\undefined}
}%
\providecommand \@ifnum [1]{%
 \ifnum #1\expandafter \@firstoftwo
 \else \expandafter \@secondoftwo
 \fi
}%
\providecommand \@ifx [1]{%
 \ifx #1\expandafter \@firstoftwo
 \else \expandafter \@secondoftwo
 \fi
}%
\providecommand \natexlab [1]{#1}%
\providecommand \enquote  [1]{``#1''}%
\providecommand \bibnamefont  [1]{#1}%
\providecommand \bibfnamefont [1]{#1}%
\providecommand \citenamefont [1]{#1}%
\providecommand \href@noop [0]{\@secondoftwo}%
\providecommand \href [0]{\begingroup \@sanitize@url \@href}%
\providecommand \@href[1]{\@@startlink{#1}\@@href}%
\providecommand \@@href[1]{\endgroup#1\@@endlink}%
\providecommand \@sanitize@url [0]{\catcode `\\12\catcode `\$12\catcode
  `\&12\catcode `\#12\catcode `\^12\catcode `\_12\catcode `\%12\relax}%
\providecommand \@@startlink[1]{}%
\providecommand \@@endlink[0]{}%
\providecommand \url  [0]{\begingroup\@sanitize@url \@url }%
\providecommand \@url [1]{\endgroup\@href {#1}{\urlprefix }}%
\providecommand \urlprefix  [0]{URL }%
\providecommand \Eprint [0]{\href }%
\providecommand \doibase [0]{http://dx.doi.org/}%
\providecommand \selectlanguage [0]{\@gobble}%
\providecommand \bibinfo  [0]{\@secondoftwo}%
\providecommand \bibfield  [0]{\@secondoftwo}%
\providecommand \translation [1]{[#1]}%
\providecommand \BibitemOpen [0]{}%
\providecommand \bibitemStop [0]{}%
\providecommand \bibitemNoStop [0]{.\EOS\space}%
\providecommand \EOS [0]{\spacefactor3000\relax}%
\providecommand \BibitemShut  [1]{\csname bibitem#1\endcsname}%
\let\auto@bib@innerbib\@empty
\bibitem [{\citenamefont {Geim}\ and\ \citenamefont
  {Grigorieva}(2013)}]{Geim2013}%
  \BibitemOpen
  \bibfield  {author} {\bibinfo {author} {\bibfnamefont {A.~K.}\ \bibnamefont
  {Geim}}\ and\ \bibinfo {author} {\bibfnamefont {I.~V.}\ \bibnamefont
  {Grigorieva}},\ }\href@noop {} {\bibfield  {journal} {\bibinfo  {journal}
  {Nature}\ }\textbf {\bibinfo {volume} {2013}},\ \bibinfo {pages} {4191}
  (\bibinfo {year} {2013})}\BibitemShut {NoStop}%
\bibitem [{\citenamefont {Chhowalla}\ \emph {et~al.}(2013)\citenamefont
  {Chhowalla}, \citenamefont {Shin}, \citenamefont {Eda}, \citenamefont {Li},
  \citenamefont {Loh},\ and\ \citenamefont {Zhang}}]{TMDC_NatChem}%
  \BibitemOpen
  \bibfield  {author} {\bibinfo {author} {\bibfnamefont {M.}~\bibnamefont
  {Chhowalla}}, \bibinfo {author} {\bibfnamefont {H.~S.}\ \bibnamefont {Shin}},
  \bibinfo {author} {\bibfnamefont {G.}~\bibnamefont {Eda}}, \bibinfo {author}
  {\bibfnamefont {L.-J.}\ \bibnamefont {Li}}, \bibinfo {author} {\bibfnamefont
  {K.~P.}\ \bibnamefont {Loh}}, \ and\ \bibinfo {author} {\bibfnamefont
  {H.}~\bibnamefont {Zhang}},\ }\href@noop {} {\bibfield  {journal} {\bibinfo
  {journal} {Nat. Chem.}\ }\textbf {\bibinfo {volume} {5}},\ \bibinfo {pages}
  {263} (\bibinfo {year} {2013})}\BibitemShut {NoStop}%
\bibitem [{\citenamefont {Bhimanapati}\ \emph {et~al.}(2015)\citenamefont
  {Bhimanapati}, \citenamefont {Lin}, \citenamefont {Meunier}, \citenamefont
  {Jung}, \citenamefont {Cha}, \citenamefont {Das}, \citenamefont {Xiao},
  \citenamefont {Son}, \citenamefont {Strano}, \citenamefont {Cooper},
  \citenamefont {Liang}, \citenamefont {Louie}, \citenamefont {Ringe},
  \citenamefont {Zhou}, \citenamefont {Kim}, \citenamefont {Naik},
  \citenamefont {Sumpter}, \citenamefont {Terrones}, \citenamefont {Xia},
  \citenamefont {Wang}, \citenamefont {Zhu}, \citenamefont {Akinwande},
  \citenamefont {Alem}, \citenamefont {Schuller}, \citenamefont {Schaak},
  \citenamefont {Terrones},\ and\ \citenamefont
  {Robinson}}]{RecentBeyondGraphene}%
  \BibitemOpen
  \bibfield  {author} {\bibinfo {author} {\bibfnamefont {G.~R.}\ \bibnamefont
  {Bhimanapati}}, \bibinfo {author} {\bibfnamefont {Z.}~\bibnamefont {Lin}},
  \bibinfo {author} {\bibfnamefont {V.}~\bibnamefont {Meunier}}, \bibinfo
  {author} {\bibfnamefont {Y.}~\bibnamefont {Jung}}, \bibinfo {author}
  {\bibfnamefont {J.}~\bibnamefont {Cha}}, \bibinfo {author} {\bibfnamefont
  {S.}~\bibnamefont {Das}}, \bibinfo {author} {\bibfnamefont {D.}~\bibnamefont
  {Xiao}}, \bibinfo {author} {\bibfnamefont {Y.}~\bibnamefont {Son}}, \bibinfo
  {author} {\bibfnamefont {M.~S.}\ \bibnamefont {Strano}}, \bibinfo {author}
  {\bibfnamefont {V.~R.}\ \bibnamefont {Cooper}}, \bibinfo {author}
  {\bibfnamefont {L.}~\bibnamefont {Liang}}, \bibinfo {author} {\bibfnamefont
  {S.~G.}\ \bibnamefont {Louie}}, \bibinfo {author} {\bibfnamefont
  {E.}~\bibnamefont {Ringe}}, \bibinfo {author} {\bibfnamefont
  {W.}~\bibnamefont {Zhou}}, \bibinfo {author} {\bibfnamefont {S.~S.}\
  \bibnamefont {Kim}}, \bibinfo {author} {\bibfnamefont {R.~R.}\ \bibnamefont
  {Naik}}, \bibinfo {author} {\bibfnamefont {B.~G.}\ \bibnamefont {Sumpter}},
  \bibinfo {author} {\bibfnamefont {H.}~\bibnamefont {Terrones}}, \bibinfo
  {author} {\bibfnamefont {F.}~\bibnamefont {Xia}}, \bibinfo {author}
  {\bibfnamefont {Y.}~\bibnamefont {Wang}}, \bibinfo {author} {\bibfnamefont
  {J.}~\bibnamefont {Zhu}}, \bibinfo {author} {\bibfnamefont {D.}~\bibnamefont
  {Akinwande}}, \bibinfo {author} {\bibfnamefont {N.}~\bibnamefont {Alem}},
  \bibinfo {author} {\bibfnamefont {J.~A.}\ \bibnamefont {Schuller}}, \bibinfo
  {author} {\bibfnamefont {R.~E.}\ \bibnamefont {Schaak}}, \bibinfo {author}
  {\bibfnamefont {M.}~\bibnamefont {Terrones}}, \ and\ \bibinfo {author}
  {\bibfnamefont {J.~A.}\ \bibnamefont {Robinson}},\ }\href@noop {} {\bibfield
  {journal} {\bibinfo  {journal} {ACS Nano}\ }\textbf {\bibinfo {volume}
  {10.1021/acsnano.5b05556}},\ \bibinfo {pages} {article asap} (\bibinfo {year}
  {2015})}\BibitemShut {NoStop}%
\bibitem [{\citenamefont {Butler}\ \emph {et~al.}(2013)\citenamefont {Butler},
  \citenamefont {Hollen}, \citenamefont {Cao}, \citenamefont {Cui},
  \citenamefont {Gupta}, \citenamefont {Guti\'{e}rrez}, \citenamefont {Heinz},
  \citenamefont {Hong}, \citenamefont {Huang}, \citenamefont {Ismach},
  \citenamefont {Johnston-Halperin}, \citenamefont {Kuno}, \citenamefont
  {Plashnitsa}, \citenamefont {Robinson}, \citenamefont {Ruoff}, \citenamefont
  {Salahuddin}, \citenamefont {Shan}, \citenamefont {Shi}, \citenamefont
  {Spencer}, \citenamefont {Terrones}, \citenamefont {Windl},\ and\
  \citenamefont {Goldberger}}]{BeyondGraphene}%
  \BibitemOpen
  \bibfield  {author} {\bibinfo {author} {\bibfnamefont {S.~Z.}\ \bibnamefont
  {Butler}}, \bibinfo {author} {\bibfnamefont {S.~M.}\ \bibnamefont {Hollen}},
  \bibinfo {author} {\bibfnamefont {L.}~\bibnamefont {Cao}}, \bibinfo {author}
  {\bibfnamefont {Y.}~\bibnamefont {Cui}}, \bibinfo {author} {\bibfnamefont
  {J.~A.}\ \bibnamefont {Gupta}}, \bibinfo {author} {\bibfnamefont {H.~R.}\
  \bibnamefont {Guti\'{e}rrez}}, \bibinfo {author} {\bibfnamefont {T.~F.}\
  \bibnamefont {Heinz}}, \bibinfo {author} {\bibfnamefont {S.~S.}\ \bibnamefont
  {Hong}}, \bibinfo {author} {\bibfnamefont {J.}~\bibnamefont {Huang}},
  \bibinfo {author} {\bibfnamefont {A.~F.}\ \bibnamefont {Ismach}}, \bibinfo
  {author} {\bibfnamefont {E.}~\bibnamefont {Johnston-Halperin}}, \bibinfo
  {author} {\bibfnamefont {M.}~\bibnamefont {Kuno}}, \bibinfo {author}
  {\bibfnamefont {V.~V.}\ \bibnamefont {Plashnitsa}}, \bibinfo {author}
  {\bibfnamefont {R.~D.}\ \bibnamefont {Robinson}}, \bibinfo {author}
  {\bibfnamefont {R.~S.}\ \bibnamefont {Ruoff}}, \bibinfo {author}
  {\bibfnamefont {S.}~\bibnamefont {Salahuddin}}, \bibinfo {author}
  {\bibfnamefont {J.}~\bibnamefont {Shan}}, \bibinfo {author} {\bibfnamefont
  {L.}~\bibnamefont {Shi}}, \bibinfo {author} {\bibfnamefont {M.~G.}\
  \bibnamefont {Spencer}}, \bibinfo {author} {\bibfnamefont {M.}~\bibnamefont
  {Terrones}}, \bibinfo {author} {\bibfnamefont {W.}~\bibnamefont {Windl}}, \
  and\ \bibinfo {author} {\bibfnamefont {J.~E.}\ \bibnamefont {Goldberger}},\
  }\href@noop {} {\bibfield  {journal} {\bibinfo  {journal} {ACS Nano}\
  }\textbf {\bibinfo {volume} {7}},\ \bibinfo {pages} {2898} (\bibinfo {year}
  {2013})}\BibitemShut {NoStop}%
\bibitem [{\citenamefont {Massicotte}\ \emph {et~al.}(2015)\citenamefont
  {Massicotte}, \citenamefont {Schmidt}, \citenamefont {Vialla}, \citenamefont
  {Sch{\"a}dler}, \citenamefont {Reserbat-Plantey}, \citenamefont {Watanabe},
  \citenamefont {Taniguchi}, \citenamefont {Tielrooij},\ and\ \citenamefont
  {Koppens}}]{Massicotte2015}%
  \BibitemOpen
  \bibfield  {author} {\bibinfo {author} {\bibfnamefont {M.}~\bibnamefont
  {Massicotte}}, \bibinfo {author} {\bibfnamefont {P.}~\bibnamefont {Schmidt}},
  \bibinfo {author} {\bibfnamefont {F.}~\bibnamefont {Vialla}}, \bibinfo
  {author} {\bibfnamefont {K.~G.}\ \bibnamefont {Sch{\"a}dler}}, \bibinfo
  {author} {\bibfnamefont {A.}~\bibnamefont {Reserbat-Plantey}}, \bibinfo
  {author} {\bibfnamefont {K.}~\bibnamefont {Watanabe}}, \bibinfo {author}
  {\bibfnamefont {T.}~\bibnamefont {Taniguchi}}, \bibinfo {author}
  {\bibfnamefont {K.~J.}\ \bibnamefont {Tielrooij}}, \ and\ \bibinfo {author}
  {\bibfnamefont {F.~H.~L.}\ \bibnamefont {Koppens}},\ }\href@noop {}
  {\bibfield  {journal} {\bibinfo  {journal} {Nat. Nanotechnol.}\ }\textbf
  {\bibinfo {volume} {online pub. 10.1038/nnano.2015.227}} (\bibinfo {year} {2015})}\BibitemShut
  {NoStop}%
\bibitem [{\citenamefont {Chang}\ \emph {et~al.}(2015)\citenamefont {Chang},
  \citenamefont {Tang}, \citenamefont {Feng}, \citenamefont {Li}, \citenamefont
  {Ma}, \citenamefont {Duan}, \citenamefont {He},\ and\ \citenamefont
  {Xue}}]{Chang2015}%
  \BibitemOpen
  \bibfield  {author} {\bibinfo {author} {\bibfnamefont {C.-Z.}\ \bibnamefont
  {Chang}}, \bibinfo {author} {\bibfnamefont {P.}~\bibnamefont {Tang}},
  \bibinfo {author} {\bibfnamefont {X.}~\bibnamefont {Feng}}, \bibinfo {author}
  {\bibfnamefont {K.}~\bibnamefont {Li}}, \bibinfo {author} {\bibfnamefont
  {X.-C.}\ \bibnamefont {Ma}}, \bibinfo {author} {\bibfnamefont
  {W.}~\bibnamefont {Duan}}, \bibinfo {author} {\bibfnamefont {K.}~\bibnamefont
  {He}}, \ and\ \bibinfo {author} {\bibfnamefont {Q.-K.}\ \bibnamefont {Xue}},\
  }\href {\doibase 10.1103/PhysRevLett.115.136801} {\bibfield  {journal}
  {\bibinfo  {journal} {Phys. Rev. Lett.}\ }\textbf {\bibinfo {volume} {115}},\
  \bibinfo {pages} {136801} (\bibinfo {year} {2015})}\BibitemShut {NoStop}%
\bibitem [{\citenamefont {Swartz}\ \emph {et~al.}(2012)\citenamefont {Swartz},
  \citenamefont {Odenthal}, \citenamefont {Hao}, \citenamefont {Ruoff},\ and\
  \citenamefont {Kawakami}}]{Swartz2012}%
  \BibitemOpen
  \bibfield  {author} {\bibinfo {author} {\bibfnamefont {A.~G.}\ \bibnamefont
  {Swartz}}, \bibinfo {author} {\bibfnamefont {P.~M.}\ \bibnamefont
  {Odenthal}}, \bibinfo {author} {\bibfnamefont {Y.}~\bibnamefont {Hao}},
  \bibinfo {author} {\bibfnamefont {R.~S.}\ \bibnamefont {Ruoff}}, \ and\
  \bibinfo {author} {\bibfnamefont {R.~K.}\ \bibnamefont {Kawakami}},\ }\href
  {\doibase 10.1021/nn303771f} {\bibfield  {journal} {\bibinfo  {journal} {ACS
  Nano}\ }\textbf {\bibinfo {volume} {6}},\ \bibinfo {pages} {10063} (\bibinfo
  {year} {2012})}\BibitemShut {NoStop}%
\bibitem [{\citenamefont {Yang}\ \emph {et~al.}(2013)\citenamefont {Yang},
  \citenamefont {Hallal}, \citenamefont {Terrade}, \citenamefont {Waintal},
  \citenamefont {Roche},\ and\ \citenamefont {Chshiev}}]{Yan2013}%
  \BibitemOpen
  \bibfield  {author} {\bibinfo {author} {\bibfnamefont {H.~X.}\ \bibnamefont
  {Yang}}, \bibinfo {author} {\bibfnamefont {A.}~\bibnamefont {Hallal}},
  \bibinfo {author} {\bibfnamefont {D.}~\bibnamefont {Terrade}}, \bibinfo
  {author} {\bibfnamefont {X.}~\bibnamefont {Waintal}}, \bibinfo {author}
  {\bibfnamefont {S.}~\bibnamefont {Roche}}, \ and\ \bibinfo {author}
  {\bibfnamefont {M.}~\bibnamefont {Chshiev}},\ }\href {\doibase
  10.1103/PhysRevLett.110.046603} {\bibfield  {journal} {\bibinfo  {journal}
  {Phys. Rev. Lett.}\ }\textbf {\bibinfo {volume} {110}},\ \bibinfo {pages}
  {046603} (\bibinfo {year} {2013})}\BibitemShut {NoStop}%
\bibitem [{\citenamefont {Yu}\ \emph {et~al.}(2011)\citenamefont {Yu},
  \citenamefont {Kanazawa}, \citenamefont {Onose}, \citenamefont {Kimoto},
  \citenamefont {Zhang}, \citenamefont {Ishiwata}, \citenamefont {Matsui},\
  and\ \citenamefont {Tokura}}]{Yu2011}%
  \BibitemOpen
  \bibfield  {author} {\bibinfo {author} {\bibfnamefont {X.~Z.}\ \bibnamefont
  {Yu}}, \bibinfo {author} {\bibfnamefont {N.}~\bibnamefont {Kanazawa}},
  \bibinfo {author} {\bibfnamefont {Y.}~\bibnamefont {Onose}}, \bibinfo
  {author} {\bibfnamefont {K.}~\bibnamefont {Kimoto}}, \bibinfo {author}
  {\bibfnamefont {W.~Z.}\ \bibnamefont {Zhang}}, \bibinfo {author}
  {\bibfnamefont {S.}~\bibnamefont {Ishiwata}}, \bibinfo {author}
  {\bibfnamefont {Y.}~\bibnamefont {Matsui}}, \ and\ \bibinfo {author}
  {\bibfnamefont {Y.}~\bibnamefont {Tokura}},\ }\href@noop {} {\bibfield
  {journal} {\bibinfo  {journal} {Nat. Mat.}\ }\textbf {\bibinfo {volume}
  {10}},\ \bibinfo {pages} {106} (\bibinfo {year} {2011})}\BibitemShut
  {NoStop}%
\bibitem [{\citenamefont {Banerjee}\ \emph {et~al.}(2014)\citenamefont
  {Banerjee}, \citenamefont {Rowland}, \citenamefont {Erten},\ and\
  \citenamefont {Randeria}}]{Banerjee2014}%
  \BibitemOpen
  \bibfield  {author} {\bibinfo {author} {\bibfnamefont {S.}~\bibnamefont
  {Banerjee}}, \bibinfo {author} {\bibfnamefont {J.}~\bibnamefont {Rowland}},
  \bibinfo {author} {\bibfnamefont {O.}~\bibnamefont {Erten}}, \ and\ \bibinfo
  {author} {\bibfnamefont {M.}~\bibnamefont {Randeria}},\ }\href {\doibase
  10.1103/PhysRevX.4.031045} {\bibfield  {journal} {\bibinfo  {journal} {Phys.
  Rev. X}\ }\textbf {\bibinfo {volume} {4}},\ \bibinfo {pages} {031045}
  (\bibinfo {year} {2014})}\BibitemShut {NoStop}%
\bibitem [{\citenamefont {Casto}\ \emph {et~al.}(2015)\citenamefont {Casto},
  \citenamefont {Clune}, \citenamefont {Yokosuk}, \citenamefont {Musfeldt},
  \citenamefont {Williams}, \citenamefont {Zhuang}, \citenamefont {Lin},
  \citenamefont {Xiao}, \citenamefont {Hennig}, \citenamefont {Sales},
  \citenamefont {Yan},\ and\ \citenamefont {Mandrus}}]{Casto2015}%
  \BibitemOpen
  \bibfield  {author} {\bibinfo {author} {\bibfnamefont {L.~D.}\ \bibnamefont
  {Casto}}, \bibinfo {author} {\bibfnamefont {A.~J.}\ \bibnamefont {Clune}},
  \bibinfo {author} {\bibfnamefont {M.~O.}\ \bibnamefont {Yokosuk}}, \bibinfo
  {author} {\bibfnamefont {J.~L.}\ \bibnamefont {Musfeldt}}, \bibinfo {author}
  {\bibfnamefont {T.~J.}\ \bibnamefont {Williams}}, \bibinfo {author}
  {\bibfnamefont {H.~L.}\ \bibnamefont {Zhuang}}, \bibinfo {author}
  {\bibfnamefont {M.-W.}\ \bibnamefont {Lin}}, \bibinfo {author} {\bibfnamefont
  {K.}~\bibnamefont {Xiao}}, \bibinfo {author} {\bibfnamefont {R.~G.}\
  \bibnamefont {Hennig}}, \bibinfo {author} {\bibfnamefont {B.~C.}\
  \bibnamefont {Sales}}, \bibinfo {author} {\bibfnamefont {J.-Q.}\ \bibnamefont
  {Yan}}, \ and\ \bibinfo {author} {\bibfnamefont {D.}~\bibnamefont
  {Mandrus}},\ }\href {\doibase http://dx.doi.org/10.1063/1.4914134} {\bibfield
   {journal} {\bibinfo  {journal} {APL Materials}\ }\textbf {\bibinfo {volume}
  {3}},\ \bibinfo {eid} {041515} (\bibinfo {year} {2015})}\BibitemShut
  {NoStop}%
\bibitem [{\citenamefont {Sivadas}\ \emph {et~al.}(2015)\citenamefont
  {Sivadas}, \citenamefont {Daniels}, \citenamefont {Swendsen}, \citenamefont
  {Okamoto},\ and\ \citenamefont {Xiao}}]{Sivadas2015}%
  \BibitemOpen
  \bibfield  {author} {\bibinfo {author} {\bibfnamefont {N.}~\bibnamefont
  {Sivadas}}, \bibinfo {author} {\bibfnamefont {M.~W.}\ \bibnamefont
  {Daniels}}, \bibinfo {author} {\bibfnamefont {R.~H.}\ \bibnamefont
  {Swendsen}}, \bibinfo {author} {\bibfnamefont {S.}~\bibnamefont {Okamoto}}, \
  and\ \bibinfo {author} {\bibfnamefont {D.}~\bibnamefont {Xiao}},\ }\href
  {\doibase 10.1103/PhysRevB.91.235425} {\bibfield  {journal} {\bibinfo
  {journal} {Phys. Rev. B}\ }\textbf {\bibinfo {volume} {91}},\ \bibinfo
  {pages} {235425} (\bibinfo {year} {2015})}\BibitemShut {NoStop}%
\bibitem [{\citenamefont {Williams}\ \emph {et~al.}(2015)\citenamefont
  {Williams}, \citenamefont {Aczel}, \citenamefont {Lumsden}, \citenamefont
  {Nagler}, \citenamefont {Stone}, \citenamefont {Yan},\ and\ \citenamefont
  {Mandrus}}]{William2015}%
  \BibitemOpen
  \bibfield  {author} {\bibinfo {author} {\bibfnamefont {T.~J.}\ \bibnamefont
  {Williams}}, \bibinfo {author} {\bibfnamefont {A.~A.}\ \bibnamefont {Aczel}},
  \bibinfo {author} {\bibfnamefont {M.~D.}\ \bibnamefont {Lumsden}}, \bibinfo
  {author} {\bibfnamefont {S.~E.}\ \bibnamefont {Nagler}}, \bibinfo {author}
  {\bibfnamefont {M.~B.}\ \bibnamefont {Stone}}, \bibinfo {author}
  {\bibfnamefont {J.-Q.}\ \bibnamefont {Yan}}, \ and\ \bibinfo {author}
  {\bibfnamefont {D.}~\bibnamefont {Mandrus}},\ }\href {\doibase
  10.1103/PhysRevB.92.144404} {\bibfield  {journal} {\bibinfo  {journal} {Phys.
  Rev. B}\ }\textbf {\bibinfo {volume} {92}},\ \bibinfo {pages} {144404}
  (\bibinfo {year} {2015})}\BibitemShut {NoStop}%
\bibitem [{\citenamefont {McGuire}\ \emph {et~al.}(2015)\citenamefont
  {McGuire}, \citenamefont {Dixit}, \citenamefont {Cooper},\ and\ \citenamefont
  {Sales}}]{McGuire2015}%
  \BibitemOpen
  \bibfield  {author} {\bibinfo {author} {\bibfnamefont {M.~A.}\ \bibnamefont
  {McGuire}}, \bibinfo {author} {\bibfnamefont {H.}~\bibnamefont {Dixit}},
  \bibinfo {author} {\bibfnamefont {V.~R.}\ \bibnamefont {Cooper}}, \ and\
  \bibinfo {author} {\bibfnamefont {B.~C.}\ \bibnamefont {Sales}},\ }\href
  {\doibase 10.1021/cm504242t} {\bibfield  {journal} {\bibinfo  {journal}
  {Chem. Mater.}\ }\textbf {\bibinfo {volume} {27}},\ \bibinfo {pages} {612}
  (\bibinfo {year} {2015})}\BibitemShut {NoStop}%
\bibitem [{\citenamefont {Zhang}\ \emph {et~al.}(2015)\citenamefont {Zhang},
  \citenamefont {Qu}, \citenamefont {Zhu},\ and\ \citenamefont
  {Lam}}]{Zhang2015}%
  \BibitemOpen
  \bibfield  {author} {\bibinfo {author} {\bibfnamefont {W.-B.}\ \bibnamefont
  {Zhang}}, \bibinfo {author} {\bibfnamefont {Q.}~\bibnamefont {Qu}}, \bibinfo
  {author} {\bibfnamefont {P.}~\bibnamefont {Zhu}}, \ and\ \bibinfo {author}
  {\bibfnamefont {C.-H.}\ \bibnamefont {Lam}},\ }\href
  {http://arxiv.org/abs/1507.07275} {\bibfield  {journal} {\bibinfo  {journal}
  {arXiv:1507.07275}\ } (\bibinfo {year} {2015})}\BibitemShut {NoStop}%
\bibitem [{\citenamefont {Carteaux}\ \emph {et~al.}(1995)\citenamefont
  {Carteaux}, \citenamefont {Moussa},\ and\ \citenamefont
  {Spiesser}}]{Carteaux1995}%
  \BibitemOpen
  \bibfield  {author} {\bibinfo {author} {\bibfnamefont {V.}~\bibnamefont
  {Carteaux}}, \bibinfo {author} {\bibfnamefont {F.}~\bibnamefont {Moussa}}, \
  and\ \bibinfo {author} {\bibfnamefont {M.}~\bibnamefont {Spiesser}},\
  }\href@noop {} {\bibfield  {journal} {\bibinfo  {journal} {Europhys. Lett.}\
  }\textbf {\bibinfo {volume} {29}},\ \bibinfo {pages} {251} (\bibinfo {year}
  {1995})}\BibitemShut {NoStop}%
\bibitem [{\citenamefont {Lin}\ \emph {et~al.}(2016)\citenamefont {Lin},
  \citenamefont {Zhuang}, \citenamefont {Yan}, \citenamefont {Ward},
  \citenamefont {Puretzky}, \citenamefont {Rouleau}, \citenamefont {Gai},
  \citenamefont {Liang}, \citenamefont {Meunier}, \citenamefont {Sumpter},
  \citenamefont {Ganesh}, \citenamefont {Kent}, \citenamefont {Geohegan},
  \citenamefont {Mandrus},\ and\ \citenamefont {Xiao}}]{MingWei2015}%
  \BibitemOpen
  \bibfield  {author} {\bibinfo {author} {\bibfnamefont {M.-W.}\ \bibnamefont
  {Lin}}, \bibinfo {author} {\bibfnamefont {H.~L.}\ \bibnamefont {Zhuang}},
  \bibinfo {author} {\bibfnamefont {J.}~\bibnamefont {Yan}}, \bibinfo {author}
  {\bibfnamefont {T.~Z.}\ \bibnamefont {Ward}}, \bibinfo {author}
  {\bibfnamefont {A.~A.}\ \bibnamefont {Puretzky}}, \bibinfo {author}
  {\bibfnamefont {C.~M.}\ \bibnamefont {Rouleau}}, \bibinfo {author}
  {\bibfnamefont {Z.}~\bibnamefont {Gai}}, \bibinfo {author} {\bibfnamefont
  {L.}~\bibnamefont {Liang}}, \bibinfo {author} {\bibfnamefont
  {V.}~\bibnamefont {Meunier}}, \bibinfo {author} {\bibfnamefont {B.~G.}\
  \bibnamefont {Sumpter}}, \bibinfo {author} {\bibfnamefont {P.}~\bibnamefont
  {Ganesh}}, \bibinfo {author} {\bibfnamefont {P.~R.~C.}\ \bibnamefont {Kent}},
  \bibinfo {author} {\bibfnamefont {D.~B.}\ \bibnamefont {Geohegan}}, \bibinfo
  {author} {\bibfnamefont {D.~G.}\ \bibnamefont {Mandrus}}, \ and\ \bibinfo
  {author} {\bibfnamefont {K.}~\bibnamefont {Xiao}},\ }\href {\doibase
  10.1039/C5TC03463A} {\bibfield  {journal} {\bibinfo  {journal} {J. Mater.
  Chem. C}\ ,\ \bibinfo {pages} {online pub. 10.1039/c5tc03463a}} (\bibinfo {year} {2015})}\BibitemShut
  {NoStop}%
\bibitem [{\citenamefont {Deiseroth}\ \emph {et~al.}(2006)\citenamefont
  {Deiseroth}, \citenamefont {Aleksandrov}, \citenamefont {Reiner},
  \citenamefont {Kienle},\ and\ \citenamefont {Kremer}}]{Deiseroth2006}%
  \BibitemOpen
  \bibfield  {author} {\bibinfo {author} {\bibfnamefont {H.-J.}\ \bibnamefont
  {Deiseroth}}, \bibinfo {author} {\bibfnamefont {K.}~\bibnamefont
  {Aleksandrov}}, \bibinfo {author} {\bibfnamefont {C.}~\bibnamefont {Reiner}},
  \bibinfo {author} {\bibfnamefont {L.}~\bibnamefont {Kienle}}, \ and\ \bibinfo
  {author} {\bibfnamefont {R.~K.}\ \bibnamefont {Kremer}},\ }\href@noop {}
  {\bibfield  {journal} {\bibinfo  {journal} {Eur. J. Inorg. Chem.}\ }\textbf
  {\bibinfo {volume} {2006}},\ \bibinfo {pages} {1561} (\bibinfo {year}
  {2006})}\BibitemShut {NoStop}%
\bibitem [{\citenamefont {Kanematsu}(1965)}]{Kanematsu1965}%
  \BibitemOpen
  \bibfield  {author} {\bibinfo {author} {\bibfnamefont {K.}~\bibnamefont
  {Kanematsu}},\ }\href@noop {} {\bibfield  {journal} {\bibinfo  {journal} {J.
  Phys. Soc. Jap.}\ }\textbf {\bibinfo {volume} {20}},\ \bibinfo {pages} {36}
  (\bibinfo {year} {1965})}\BibitemShut {NoStop}%
\bibitem [{\citenamefont {Chen}\ \emph {et~al.}(2013)\citenamefont {Chen},
  \citenamefont {Yang}, \citenamefont {Wang}, \citenamefont {Imai},
  \citenamefont {Ohta}, \citenamefont {Michioka}, \citenamefont {Yoshimura},\
  and\ \citenamefont {Fang}}]{Chen2013}%
  \BibitemOpen
  \bibfield  {author} {\bibinfo {author} {\bibfnamefont {B.}~\bibnamefont
  {Chen}}, \bibinfo {author} {\bibfnamefont {J.-H.}\ \bibnamefont {Yang}},
  \bibinfo {author} {\bibfnamefont {H.-D.}\ \bibnamefont {Wang}}, \bibinfo
  {author} {\bibfnamefont {M.}~\bibnamefont {Imai}}, \bibinfo {author}
  {\bibfnamefont {H.}~\bibnamefont {Ohta}}, \bibinfo {author} {\bibfnamefont
  {C.}~\bibnamefont {Michioka}}, \bibinfo {author} {\bibfnamefont
  {K.}~\bibnamefont {Yoshimura}}, \ and\ \bibinfo {author} {\bibfnamefont
  {M.-H.}\ \bibnamefont {Fang}},\ }\href@noop {} {\bibfield  {journal}
  {\bibinfo  {journal} {J. Phys. Soc. Jap.}\ }\textbf {\bibinfo {volume}
  {82}},\ \bibinfo {pages} {124711} (\bibinfo {year} {2013})}\BibitemShut
  {NoStop}%
\bibitem [{\citenamefont {Verchenko}\ \emph {et~al.}(2015)\citenamefont
  {Verchenko}, \citenamefont {Tsirlin}, \citenamefont {Sobolev}, \citenamefont
  {Presniakov},\ and\ \citenamefont {Shevelkov}}]{Verchenko2015}%
  \BibitemOpen
  \bibfield  {author} {\bibinfo {author} {\bibfnamefont {V.~Y.}\ \bibnamefont
  {Verchenko}}, \bibinfo {author} {\bibfnamefont {A.~A.}\ \bibnamefont
  {Tsirlin}}, \bibinfo {author} {\bibfnamefont {A.~V.}\ \bibnamefont
  {Sobolev}}, \bibinfo {author} {\bibfnamefont {I.~A.}\ \bibnamefont
  {Presniakov}}, \ and\ \bibinfo {author} {\bibfnamefont {A.~V.}\ \bibnamefont
  {Shevelkov}},\ }\href {\doibase 10.1021/acs.inorgchem.5b01260} {\bibfield
  {journal} {\bibinfo  {journal} {Inorg. Mater.}\ }\textbf {\bibinfo {volume}
  {54}},\ \bibinfo {pages} {8598} (\bibinfo {year} {2015})}\BibitemShut
  {NoStop}%
\bibitem [{\citenamefont {Sheldrick}(2008)}]{Shelxl97}%
  \BibitemOpen
  \bibfield  {author} {\bibinfo {author} {\bibfnamefont {G.~M.}\ \bibnamefont
  {Sheldrick}},\ }\href@noop {} {\bibfield  {journal} {\bibinfo  {journal}
  {Acta Cryst.}\ }\textbf {\bibinfo {volume} {A64}},\ \bibinfo {pages} {112}
  (\bibinfo {year} {2008})}\BibitemShut {NoStop}%
\bibitem [{\citenamefont {Parthe}\ and\ \citenamefont
  {Gelato}(1984)}]{StructureTidy}%
  \BibitemOpen
  \bibfield  {author} {\bibinfo {author} {\bibfnamefont {E.}~\bibnamefont
  {Parthe}}\ and\ \bibinfo {author} {\bibfnamefont {L.~M.}\ \bibnamefont
  {Gelato}},\ }\href@noop {} {\bibfield  {journal} {\bibinfo  {journal} {Acta
  Cryst.}\ }\textbf {\bibinfo {volume} {A40}},\ \bibinfo {pages} {169}
  (\bibinfo {year} {1984})}\BibitemShut {NoStop}%
\bibitem [{\citenamefont {Spek}(2009)}]{Platon}%
  \BibitemOpen
  \bibfield  {author} {\bibinfo {author} {\bibfnamefont {A.~L.}\ \bibnamefont
  {Spek}},\ }\href@noop {} {\bibfield  {journal} {\bibinfo  {journal} {Acta
  Cryst.}\ }\textbf {\bibinfo {volume} {D65}},\ \bibinfo {pages} {148}
  (\bibinfo {year} {2009})}\BibitemShut {NoStop}%
\bibitem [{\citenamefont {Rodríguez-Carvajal}(1993)}]{FullProf}%
  \BibitemOpen
  \bibfield  {author} {\bibinfo {author} {\bibfnamefont {J.}~\bibnamefont
  {Rodríguez-Carvajal}},\ }\href@noop {} {\bibfield  {journal} {\bibinfo
  {journal} {Physica B}\ }\textbf {\bibinfo {volume} {192}},\ \bibinfo {pages}
  {55} (\bibinfo {year} {1993})}\BibitemShut {NoStop}%
\bibitem [{\citenamefont {Malaman}\ \emph {et~al.}(1980)\citenamefont
  {Malaman}, \citenamefont {Steinmetz},\ and\ \citenamefont
  {Roques}}]{Malaman1980}%
  \BibitemOpen
  \bibfield  {author} {\bibinfo {author} {\bibfnamefont {B.}~\bibnamefont
  {Malaman}}, \bibinfo {author} {\bibfnamefont {J.}~\bibnamefont {Steinmetz}},
  \ and\ \bibinfo {author} {\bibfnamefont {B.}~\bibnamefont {Roques}},\
  }\href@noop {} {\bibfield  {journal} {\bibinfo  {journal} {J. Less-Common
  Metals}\ }\textbf {\bibinfo {volume} {75}},\ \bibinfo {pages} {155} (\bibinfo
  {year} {1980})}\BibitemShut {NoStop}%
\bibitem [{\citenamefont {Katsuraki}(1964)}]{Katsuraki1964}%
  \BibitemOpen
  \bibfield  {author} {\bibinfo {author} {\bibfnamefont {H.}~\bibnamefont
  {Katsuraki}},\ }\href@noop {} {\bibfield  {journal} {\bibinfo  {journal} {J.
  Phys. Soc. Jap.}\ }\textbf {\bibinfo {volume} {19}},\ \bibinfo {pages} {863}
  (\bibinfo {year} {1964})}\BibitemShut {NoStop}%
\bibitem [{\citenamefont {Germagnoli}\ \emph {et~al.}(1966)\citenamefont
  {Germagnoli}, \citenamefont {Lamborizio}, \citenamefont {Mora},\ and\
  \citenamefont {Ortalli}}]{Germagnoli1966}%
  \BibitemOpen
  \bibfield  {author} {\bibinfo {author} {\bibfnamefont {E.}~\bibnamefont
  {Germagnoli}}, \bibinfo {author} {\bibfnamefont {C.}~\bibnamefont
  {Lamborizio}}, \bibinfo {author} {\bibfnamefont {S.}~\bibnamefont {Mora}}, \
  and\ \bibinfo {author} {\bibfnamefont {I.}~\bibnamefont {Ortalli}},\
  }\href@noop {} {\bibfield  {journal} {\bibinfo  {journal} {Il Nuovo Cimento}\
  }\textbf {\bibinfo {volume} {42B}},\ \bibinfo {pages} {314} (\bibinfo {year}
  {1966})}\BibitemShut {NoStop}%
\bibitem [{\citenamefont {Albertini}\ \emph {et~al.}(1998)\citenamefont
  {Albertini}, \citenamefont {Pareti}, \citenamefont {Deriu}, \citenamefont
  {Negri}, \citenamefont {Calestani}, \citenamefont {Moze}, \citenamefont
  {Kennedy},\ and\ \citenamefont {Sonntag}}]{Albertini1998}%
  \BibitemOpen
  \bibfield  {author} {\bibinfo {author} {\bibfnamefont {F.}~\bibnamefont
  {Albertini}}, \bibinfo {author} {\bibfnamefont {L.}~\bibnamefont {Pareti}},
  \bibinfo {author} {\bibfnamefont {A.}~\bibnamefont {Deriu}}, \bibinfo
  {author} {\bibfnamefont {D.}~\bibnamefont {Negri}}, \bibinfo {author}
  {\bibfnamefont {G.}~\bibnamefont {Calestani}}, \bibinfo {author}
  {\bibfnamefont {O.}~\bibnamefont {Moze}}, \bibinfo {author} {\bibfnamefont
  {S.~J.}\ \bibnamefont {Kennedy}}, \ and\ \bibinfo {author} {\bibfnamefont
  {R.}~\bibnamefont {Sonntag}},\ }\href@noop {} {\bibfield  {journal} {\bibinfo
   {journal} {J. Appl. Phys.}\ }\textbf {\bibinfo {volume} {84}},\ \bibinfo
  {pages} {401} (\bibinfo {year} {1998})}\BibitemShut {NoStop}%
\bibitem [{\citenamefont {Wohlfarth}(1978)}]{Wohlfarth1978}%
  \BibitemOpen
  \bibfield  {author} {\bibinfo {author} {\bibfnamefont {E.~P.}\ \bibnamefont
  {Wohlfarth}},\ }\href@noop {} {\bibfield  {journal} {\bibinfo  {journal} {J.
  Mag. Mag. Mater.}\ }\textbf {\bibinfo {volume} {7}},\ \bibinfo {pages} {113}
  (\bibinfo {year} {1978})}\BibitemShut {NoStop}%
\bibitem [{\citenamefont {Moriya}(1979)}]{Moriya1979}%
  \BibitemOpen
  \bibfield  {author} {\bibinfo {author} {\bibfnamefont {T.}~\bibnamefont
  {Moriya}},\ }\href@noop {} {\bibfield  {journal} {\bibinfo  {journal} {J.
  Mag. Mag. Mater.}\ }\textbf {\bibinfo {volume} {14}},\ \bibinfo {pages} {1}
  (\bibinfo {year} {1979})}\BibitemShut {NoStop}%
\bibitem [{Note1()}]{Note1}%
  \BibitemOpen
  \bibinfo {note} {See Supplemental Material at [URL will be inserted by
  publisher] for powder neutron diffraction data, anomalous Hall effect data,
  electrical resistivity, thermal transport, and specific heat
  data.}\BibitemShut {Stop}%
\end{thebibliography}

\begin{thebibliography}{1}%
\makeatletter
\providecommand \@ifxundefined [1]{%
 \@ifx{#1\undefined}
}%
\providecommand \@ifnum [1]{%
 \ifnum #1\expandafter \@firstoftwo
 \else \expandafter \@secondoftwo
 \fi
}%
\providecommand \@ifx [1]{%
 \ifx #1\expandafter \@firstoftwo
 \else \expandafter \@secondoftwo
 \fi
}%
\providecommand \natexlab [1]{#1}%
\providecommand \enquote  [1]{``#1''}%
\providecommand \bibnamefont  [1]{#1}%
\providecommand \bibfnamefont [1]{#1}%
\providecommand \citenamefont [1]{#1}%
\providecommand \href@noop [0]{\@secondoftwo}%
\providecommand \href [0]{\begingroup \@sanitize@url \@href}%
\providecommand \@href[1]{\@@startlink{#1}\@@href}%
\providecommand \@@href[1]{\endgroup#1\@@endlink}%
\providecommand \@sanitize@url [0]{\catcode `\\12\catcode `\$12\catcode
  `\&12\catcode `\#12\catcode `\^12\catcode `\_12\catcode `\%12\relax}%
\providecommand \@@startlink[1]{}%
\providecommand \@@endlink[0]{}%
\providecommand \url  [0]{\begingroup\@sanitize@url \@url }%
\providecommand \@url [1]{\endgroup\@href {#1}{\urlprefix }}%
\providecommand \urlprefix  [0]{URL }%
\providecommand \Eprint [0]{\href }%
\providecommand \doibase [0]{http://dx.doi.org/}%
\providecommand \selectlanguage [0]{\@gobble}%
\providecommand \bibinfo  [0]{\@secondoftwo}%
\providecommand \bibfield  [0]{\@secondoftwo}%
\providecommand \translation [1]{[#1]}%
\providecommand \BibitemOpen [0]{}%
\providecommand \bibitemStop [0]{}%
\providecommand \bibitemNoStop [0]{.\EOS\space}%
\providecommand \EOS [0]{\spacefactor3000\relax}%
\providecommand \BibitemShut  [1]{\csname bibitem#1\endcsname}%
\let\auto@bib@innerbib\@empty
\bibitem [{\citenamefont {Chen}\ \emph {et~al.}(2013)\citenamefont {Chen},
  \citenamefont {Yang}, \citenamefont {Wang}, \citenamefont {Imai},
  \citenamefont {Ohta}, \citenamefont {Michioka}, \citenamefont {Yoshimura},\
  and\ \citenamefont {Fang}}]{Chen2013}%
  \BibitemOpen
  \bibfield  {author} {\bibinfo {author} {\bibfnamefont {B.}~\bibnamefont
  {Chen}}, \bibinfo {author} {\bibfnamefont {J.-H.}\ \bibnamefont {Yang}},
  \bibinfo {author} {\bibfnamefont {H.-D.}\ \bibnamefont {Wang}}, \bibinfo
  {author} {\bibfnamefont {M.}~\bibnamefont {Imai}}, \bibinfo {author}
  {\bibfnamefont {H.}~\bibnamefont {Ohta}}, \bibinfo {author} {\bibfnamefont
  {C.}~\bibnamefont {Michioka}}, \bibinfo {author} {\bibfnamefont
  {K.}~\bibnamefont {Yoshimura}}, \ and\ \bibinfo {author} {\bibfnamefont
  {M.-H.}\ \bibnamefont {Fang}},\ }\href@noop {} {\bibfield  {journal}
  {\bibinfo  {journal} {J. Phys. Soc. Jap.}\ }\textbf {\bibinfo {volume}
  {82}},\ \bibinfo {pages} {124711} (\bibinfo {year} {2013})}\BibitemShut
  {NoStop}%
\end{thebibliography}

%

\end{document}